\documentclass[journal]{IEEEtran}

\usepackage[utf8]{inputenc}
\usepackage[T1]{fontenc}
\usepackage{amsmath,amssymb,amsthm}
\usepackage{mathtools}
\usepackage{graphicx}
\usepackage{booktabs}
\usepackage{xurl}
\usepackage{multirow}
\usepackage{array}
\usepackage{enumitem}
\usepackage[table]{xcolor}
\usepackage{float}
\usepackage{hyperref}
\usepackage{bookmark}
\usepackage[expansion=false]{microtype}
\usepackage{tikz}
\usetikzlibrary{positioning, arrows.meta, shapes.geometric, calc}

\newtheorem{theorem}{Theorem}
\newtheorem{proposition}[theorem]{Proposition}
\newtheorem{corollary}[theorem]{Corollary}
\newtheorem{lemma}[theorem]{Lemma}
\newtheorem{definition}[theorem]{Definition}
\newtheorem{remark}[theorem]{Remark}

\DeclareMathOperator*{\argmax}{arg\,max}
\DeclareMathOperator{\Var}{Var}
\DeclareMathOperator{\CV}{CV}

\newcommand{\NO}{\mathrm{NO}}
\newcommand{\MoS}{\mathrm{MoS}}
\newcommand{\PoD}{\mathrm{PoD}}
\newcommand{\PoS}{\mathrm{PoS}}
\newcommand{\iid}{\stackrel{\text{i.i.d.}}{\sim}}
\newif\ifshowchanges
\showchangesfalse
\newcommand{\rev}[2][]{\begingroup\ifshowchanges\color{blue}\fi#2\endgroup}
\ifdefined\ShowMinorChanges
\renewcommand{\rev}[2][]{\begingroup
  \def\currenttag{#1}\def\tagA{MR.1}\def\tagB{MR.2}%
  \def\tagC{ED.1}\def\tagD{ED.3}\def\tagE{AI}%
  \ifx\currenttag\tagA\color{blue}\fi
  \ifx\currenttag\tagB\color{blue}\fi
  \ifx\currenttag\tagC\color{blue}\fi
  \ifx\currenttag\tagD\color{blue}\fi
  \ifx\currenttag\tagE\color{blue}\fi#2\endgroup}
\fi

\title{Dynamic Coalition Formation Under Capacity\\Constraints: A Regime-Switching Value-Investing Game}
\author{Rosario Patan\`e\\SAMOVAR, T\'el\'ecom SudParis, Institut Polytechnique de Paris, Palaiseau, France\\rosario.patane@telecom-sudparis.eu}

\begin{document}
\maketitle

\begin{abstract}
Edge computing service deployment requires the Network Operator (NO)
and Service Providers (SPs) to coordinate investment and resource allocation
under uncertain demand. This paper develops an audit-based service
management framework for daily coalition formation with fixed capacity
and service-level agreement (SLA) reservations. A two-stage coalitional
game separates infrastructure investment from operational participation.
Demand follows a Geometric Random Walk (GRW), and SPs apply an ordered
admission and suspension rule using reported marginal contributions.
With the NO held fixed, the operational coalition value is submodular;
mandatory reservations can produce negative Shapley payoffs.
We derive a separable benchmark, compute operational Shapley allocations,
and give a sufficient condition for convergence to an individual equilibrium.
A probabilistic audit mechanism provides a conditional daily truthfulness
guarantee under verified settlement and enforceable penalties.
For five service classes over three years, the reference run gains 12.4\%
over static grand-coalition membership and recovers 94.6\% of the
full-information optimum. Across 30 runs, the median gain is 13.3\%.
Sensitivity tests examine policy parameters, capacity and alternative
demand models. The gain depends on capacity and update order;
abundant capacity can make the admission rule underperform static membership.
\end{abstract}


\section{Introduction}
\label{sec:intro}
Managing multi-tenant edge services requires solving jointly resource
allocation, service admission, and Service-Level Agreement (SLA)
enforcement under shared capacity. Edge Computing (EC) brings compute
resources close to users, enabling latency-critical applications such
as autonomous driving, augmented/virtual reality (AR/VR),
cloud gaming, and the Internet of Things (IoT)
\cite{satyanarayanan2017}. Although \rev[R4.m1]{International Data Corporation (IDC) projected} worldwide EC spending of
\$232 billion for 2024~\cite{idc2024}, large-scale deployment
remains limited because of an economic rather than technological
bottleneck. The Network Operator (NO) bears the infrastructure capital expenditure (CAPEX),
whereas Service Providers (SPs) capture most service revenues, leaving
neither side with sufficient incentive to invest independently.
Coalitional co-investment addresses this misalignment by allowing the
NO and SPs to share costs and revenues. A foundational
model~\cite{patane2023ccnc} showed that the resulting cooperative game
is convex and that the Shapley value aligns incentives, assuming
deterministic demand, static coalition membership, and fully informed
players. Three assumptions limit current co-investment formulations.

\paragraph{(A1) Deterministic demand}
Existing models assume known traffic and revenue trajectories
\cite{patane2023ccnc}. Recent stochastic extensions account for revenue
uncertainty~\cite{sakr2025icc}, but still rely on fixed coalition
membership.

\paragraph{(A2) Static membership}
Classical co-investment fixes coalition membership at deployment.
Dynamic extensions~\cite{sakr2025dynamic} introduce entry and exit
fees, but coalition evolution remains centrally managed rather than
driven by decentralized daily decisions.

\paragraph{(A3) Full information}
Existing formulations assume fully informed, rational players. In
practice, SPs observe only local information and must make decisions
under uncertainty. Graham's margin-of-safety (MoS) principle
\cite{graham1949} provides a natural behavioural rule for such
settings.

\paragraph{Proposed framework}
We address (A1)--(A3) jointly with a decentralized value-investing
framework that separates the irreversible deployment decision from
daily coalition formation, and that admits or revokes tenants on a
margin-of-safety criterion motivated by residual-horizon traffic uncertainty. Our contributions are as
follows.
\begin{enumerate}[nosep,leftmargin=*]
\item We formulate a two-stage coalitional game with stochastic
Geometric Random Walk (GRW) demand that decouples long-term capacity
deployment from daily coalition formation.
\item We establish submodularity conditional on the NO's
presence and characterize Shapley allocations under capacity and
SLA constraints, including the possibility of negative operational payoffs.
\item We design a decentralized value-investing admission rule with a
volatility-scaled margin of safety and establish a finite bound
for detecting repeated states of the ordered update rule.
\item We develop an audit-and-trust mechanism that provides a
conditional daily truthfulness guarantee for the Network Operator
and characterize the audit frequency minimizing the stated criterion.
\item Numerical experiments over five edge service classes achieve
Price of Decentralisation (PoD) $=1.057$ and
dynamic/static value ratio (PoS) $=1.124$: a $+12.4\%$ gain over the static
grand coalition, evaluated with the same operational capacity model.
\end{enumerate}

The remainder of the paper is organised as follows. Section~\ref{sec:related} reviews related work and
Section~\ref{sec:model} introduces the model;
Sections~\ref{sec:analysis}--\ref{sec:sp} develop the theoretical
results and Section~\ref{sec:pod} defines the evaluation metrics;
Section~\ref{sec:results} reports the numerical evaluation and
Section~\ref{sec:concl} concludes. Proofs of the main results are in the Appendix; the acronym
glossary is in Table~\ref{tab:notation}. The complete simulation code, parameter files,
and scripts generating the figures and numerical results are provided
in the accompanying reproducibility package. The original public
repository is
\url{https://github.com/patan3saro/EdgeValueInvesting-Sim} under the
MIT licence.

\section{Related Work}
\label{sec:related}
We organise prior work along three axes that map directly to the
limitations of Section~\ref{sec:intro}: coalitional games for networks
and service management, edge-computing co-investment, and
value-investing principles in mechanism design.

\paragraph{Coalitional games for networks and resource sharing}
\rev[R4.m3]{The tutorial of Saad et al.~\cite{saad2009tutorial} presents
cooperative game theory and merge-and-split coalition formation for
communication networks. Applications include stochastic spectrum
sharing~\cite{karami2018stochastic}, cooperative cloud provisioning
under uncertain demand~\cite{niyato2011coalition}, and content
distribution in vehicular networks~\cite{wang2013pcd}.
Mashayekhy et al.~\cite{mashayekhy2015federation} study cloud
federations with a normalized Banzhaf profit-sharing rule and
merge-and-split formation. Zafari et al.~\cite{zafari2021share}, in
IEEE Transactions on Network and Service Management (TNSM), study
resource sharing in mobile edge clouds under capacity constraints.
These settings differ in their stochastic assumptions and allocation rules.
}\emph{Our position.} We retain decentralized coalition participation, but here the source of dynamism is a stochastic,
player-specific traffic process \eqref{eq:grw} that changes coalition values and capacity shadow
prices over time. The resulting participation decisions account for
the capacity displaced by mandatory SLA allocations.

\paragraph{Edge computing co-investment}
A foundational coalitional model for edge co-investment with one NO
and multiple SPs was proposed in~\cite{patane2023ccnc}: convexity of
the resulting cooperative game was established and the Shapley payoff
derived in closed form. The model is \emph{deterministic} and
\emph{static}; it motivates a static-membership comparator; our experiments
evaluate Grand under the present operational model
(Section~\ref{sec:results}). \rev[R3.m1,R4.m3]{Sakr et al.~\cite{sakr2025icc} introduce revenue uncertainty;
their journal extension~\cite{sakr2026annals} addresses dependent
revenue fluctuations and probabilistic stability and profitability.
The dynamic-participation study~\cite{sakr2025dynamic} updates
capacity and membership across decision epochs, with entry fees,
exit penalties and compensation. }The present mechanism is based on distributed SP decisions with operator-computed value signals and emerges
from a daily rule that adapts to realised demand; players exit when
their own margin-of-safety condition fails; and declarations are supported by an audit-and-trust mechanism
with explicit settlement and enforceability assumptions. \emph{Our position.} We extend this line on three axes at
once: the coalition re-forms daily under stochastic traffic rather
than at exogenous re-negotiation epochs; the join/leave decision is
decentralised and operates under partial information; and the
mechanism addresses daily NO misreporting under explicit audit assumptions, an issue none of
\cite{patane2023ccnc,sakr2025icc,sakr2025dynamic} addresses.

\paragraph{Value investing and margin of safety}
Graham's \emph{margin of safety}~\cite{graham1949} prescribes
committing capital only when intrinsic value exceeds market price by a
buffer absorbing valuation error. Threshold policies in optimal
stopping, hysteresis in repeated games, and robust mechanism
design~\cite{aghassi2006robust} are structurally related, but none
makes the analogy explicit. \emph{Our position.} We make three
connections explicit. First, SP~$i$'s marginal contribution
$\Delta_i(d)$, the operational value gained on entry or lost on
departure at day $d$, is the \emph{intrinsic value} analogue, while the
current coalition is $\mathcal S(d)$, its SP subset is
$\mathcal P(d)$, and $v_d(\mathcal S(d))$ is its daily operational
value. With $|\cdot|$ denoting cardinality, the average
$v_d(\mathcal S(d))/|\mathcal P(d)|$ is the \emph{market price} analogue.
Second, let $\theta_{\min}$ be a positive base margin,
$\kappa$ a sensitivity coefficient, $\sigma_i$ daily log-demand
volatility, $D_{\mathrm{inv}}$ the horizon in days, and $\theta_i(d)$
the join threshold. We set
$\theta_i(d)=\theta_{\min}+\kappa\sigma_i\sqrt{D_{\mathrm{inv}}-d}$.
This scaling is motivated by the small-variance approximation
of future traffic uncertainty in Proposition~\ref{prop:CV}, while
its coefficients are policy parameters. Third, the distinct leave
threshold specifies a retention criterion
(Remark~\ref{rem:asym_theta}); it does not prove hysteresis.

\section{Service Management Model and Coalitional Co-Investment}
\label{sec:model}

We model edge co-investment as a coalitional game with transferable
utility (TU)~\cite{osborne1994}. Let $v$ denote its characteristic
function. The player set
\rev[R4.M6]{\textcolor{black}{$\mathcal I=\{1,\ldots,M,\NO\}$} is the grand coalition:
\textcolor{black}{$M$} counts SPs and \textcolor{black}{$M+1$} counts all players. Calligraphic symbols
denote sets, and upright
\textcolor{black}{$\mathrm N(0,1)$} denotes the standard normal distribution.
The symbol \textcolor{black}{$\NO$} identifies the unique infrastructure owner.}

\paragraph{Two-Stage Structure}
Let $D_{\mathrm{inv}}$ be the investment horizon in days and
$d\in\{0,1,\ldots,D_{\mathrm{inv}}\}$ the day index; each day is a decision epoch. The active coalition
$\mathcal{S}(d) \subseteq \mathcal{I}$ is the subset of players
sharing resources on day~$d$, with active SPs
$\mathcal{P}(d) = \mathcal{S}(d) \setminus \{\NO\}$. Here $\setminus$
denotes set difference and removes the NO from the active set.

\paragraph{Stage~I ($d=0$)}
Let $p_i\geq0$ be player $i$'s upfront contribution in dollars and
$q_{\mathrm{cap}}>0$ the hardware cost per millicore (mc).
The investors deploy $C=q_{\mathrm{cap}}^{-1}\sum_{i\in\mathcal I}p_i$.
The CAPEX $q_{\mathrm{cap}}C$ is sunk; initially $\mathcal S(0)=\mathcal I$.
\rev[R3.m3]{Operational exit temporarily suspends service and preserves
ownership. Let \textcolor{black}{$\omega_i=p_i/(q_{\mathrm{cap}}C)$} be the ownership share,
\textcolor{black}{$\chi\in[0,1]$} the fraction reserved for investors, and \textcolor{black}{$\varphi_i(d)$}
the operational Shapley payoff, zero for an inactive SP. If \textcolor{black}{$v_d$}
denotes daily coalition value, total receipts in value units are
\textcolor{black}{$x_i(d)=\chi\omega_i v_d(\mathcal S(d))+(1-\chi)\varphi_i(d)$}.
An inactive investor retains the first term. With conversion
\textcolor{black}{$s_{\mathrm{vu}}>0$} dollars per value unit, discount factors \textcolor{black}{$b_d>0$}
and outside-option value \textcolor{black}{$o_i$} in dollars, let \textcolor{black}{$\mathbb E$} denote
expectation. Participation requires
\textcolor{black}{$s_{\mathrm{vu}}\mathbb E[\sum_d b_dx_i(d)]\geq p_i+o_i$}.
The experiments take \textcolor{black}{$C$} as fixed and report the operational account;
they do not calibrate this contract or establish its participation feasibility.}

\paragraph{Stage~II ($d \geq 1$)}
The index $i\in\{1,\ldots,M\}$ identifies an SP.
With $C$ fixed, $\mathcal{S}(d)$ re-forms each day via the
value-investing rule of Section~\ref{sec:vi}. Each day defines a
cooperative game $(\mathcal{I},v_d)$. For an SP set
$\mathcal P(\mathcal S)=\mathcal S\setminus\{\NO\}$, let $h_i$ be
the compute allocation, $\underline h_i$ its contractual minimum,
both in mc, and $\Lambda_d^i$ the daily traffic scale in requests per
second (req/s). Write $h=(h_i)_{i\in\mathcal P(\mathcal S)}$ and
$\underline h=(\underline h_i)_{i\in\mathcal P(\mathcal S)}$;
$h\geq\underline h$ holds componentwise. Let $U_d^i$ be the daily utility.
For feasible coalitions containing the NO, the coalition value is
\begin{equation}
  v_d(\mathcal{S}) = \max_{h \,\geq\, \underline{h}}
  \sum_{i \in \mathcal{P}(\mathcal{S})} U^i_d(h_i; \Lambda^i_d)
  \;\text{s.t.}\;
  \sum_{i \in \mathcal{P}(\mathcal{S})} h_i \leq C,
  \label{eq:char_function}
\end{equation}
The Shapley value is evaluated on the sub-coalitions of the
realized active set, including counterfactual coalitions without the NO.
The active set is produced by the ordered daily rule of
Section~\ref{sec:vi}; it need not be an equilibrium.
\rev[MR.2]{Fixed capacity represents the installed central processing
unit (CPU) budget determined in Stage~I and held constant over the
investment horizon. The framework targets long-term infrastructure
co-investment and daily service-coalition management within this
predetermined budget. General elastic orchestration, including dynamic
scaling, cloud offloading, online expansion, or joint management of
heterogeneous resources such as CPU, memory and bandwidth, lies outside
the present model.}

\begin{figure}[t]
\centering
\begin{tikzpicture}[
  stage/.style={rectangle, draw=black!80, fill=blue!10, thick,
                minimum width=2.6cm, minimum height=1cm,
                align=center, font=\small},
  process/.style={rectangle, draw=black!60, fill=gray!5,
                  minimum width=2.4cm, minimum height=1cm,
                  align=center, font=\footnotesize},
  arrow/.style={->, >=stealth, thick}
]
\node[stage] (invest) at (0,0)
  {\textbf{Stage I}\\Capacity\\$(d=0)$};
\node[process, below=0.2cm of invest]
  {Deploy $C = \tfrac{1}{q_{\mathrm{cap}}}\sum p_i$\\(sunk)};
\node[stage, right=2cm of invest] (repeat)
  {\textbf{Stage II}\\Daily formation\\$(d \geq 1)$};
\node[process, below=0.2cm of repeat]
  {Join/leave\\via value ratios};
\draw[arrow] (invest) -- (repeat)
  node[midway, above, font=\scriptsize] {$C$ fixed};
\end{tikzpicture}
\caption{Two-stage game structure. Stage~I is the irreversible
capacity decision; Stage~II is the daily coalition formation under
$(\mathcal{I}, v_d)$.}
\label{fig:two_stage}
\end{figure}
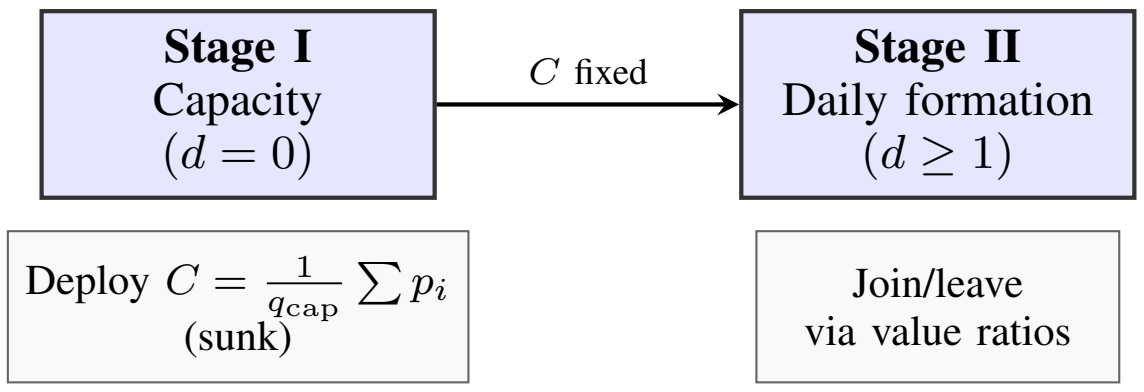

\paragraph{Traffic Model}
\label{sec:traffic}

Each day has $T = 96$ timeslots of $\Delta t = 900$\,s. The arrival
rate is $l^i_{d,t} = \Lambda^i_d \cdot f(t)$~[req/s], with $f$ the
deterministic intra-day profile~\cite{vela2016}. The slot index is $t\in\{1,\ldots,T\}$ and $\Delta t$ is slot duration.
The daily aggregate $F=\sum_{t=1}^{T}f(t)$ is a dimensionless profile sum.
Only this aggregate enters the utility, and the simulations fix
$F=96$; they do not simulate a queue within individual slots.

Let $\Lambda_0^i>0$ be initial demand, $\mu_i$ the daily drift,
and $\sigma_i$ volatility per square root of a day. The innovations
$\epsilon_d^i$ are independent and identically distributed (i.i.d.)
over SPs and days, with $\epsilon_d^i\iid\mathrm N(0,1)$.
Using one-day steps, the daily scale follows a Geometric Random Walk (GRW):
\begin{equation}
  \Lambda^i_d \;=\; \Lambda^i_{d-1} \,
  \exp\!\bigl[(\mu_i - \tfrac{\sigma_i^2}{2}) + \sigma_i \epsilon_d^i\bigr],
  \quad d\geq1,
  \label{eq:grw}
\end{equation}
Here $\mu_i$ has units day$^{-1}$ and $\sigma_i$ has units
day$^{-1/2}$. The corrective term ensures
$\mathbb E[\Lambda_d^i]=\Lambda_0^i e^{\mu_i d}$, where
$\mathbb E$ denotes expectation.

The GRW provides a positive multiplicative demand model with
tractable conditional moments. It represents scenarios of evolving
service popularity, rather than a calibrated model of every traffic
feature. Independent Gaussian increments do not capture bursts,
periodicity or long-range dependence; alternative synthetic traffic
models are evaluated in Section~\ref{sec:results}.

\paragraph{Utility, Capacity, and SLA}
\label{sec:sla}

Let $\xi>0$ be the saturation rate in mc$^{-1}$ and
$\beta_i>0$ the traffic-to-value coefficient in vu/(req/s), where
vu denotes a normalized value unit. The daily utility of SP~$i$ is
\begin{equation}
  U^i_d(h_i) \;=\; \beta_i \, \Lambda^i_d \, F \,
  \bigl(1 - e^{-\xi h_i}\bigr),
  \quad \beta_{\NO} = 0,
  \label{eq:Ud}
\end{equation}
The coefficient $\beta_{\NO}=0$ means that the NO has no direct
service revenue. Thus $\beta_i$ is not a calibrated dollar tariff per request.
A monetary interpretation would require a conversion using actual
request counts, including slot duration. \rev[R4.M2,R4.m3]{The exponential utility is a modeling choice for diminishing returns.
Salchow~\cite{salchow2008} discusses multicore scaling limitations;
that source does not derive this utility. Memory and cache effects
are outside the numerical specification.}

\begin{lemma}[Concavity]
\label{lem:concave}
$U^i_d(h_i)$ is strictly concave and strictly increasing in $h_i$ on
$\mathbb{R}_{+}$.
\end{lemma}
\begin{proof}
$\partial_h U^i_d = \beta_i \Lambda^i_d F\,\xi\,e^{-\xi h} > 0$ and
$\partial^2_h U^i_d = -\beta_i \Lambda^i_d F\,\xi^2 e^{-\xi h} < 0$
for all $h \geq 0$, since $\beta_i \Lambda^i_d F > 0$ and $\xi > 0$.
\end{proof}

The SLA reservation $\underline h_i\geq0$ is the minimum compute
allocation assigned to service $i$ by contract.
The experiments use the reservation vector
$(30,300,1500,0.3,112.5)$ mc as an input. These floors are not derived
from a queueing model and do not by themselves prove simultaneous
throughput and end-to-end latency guarantees. Autonomous driving has
the largest reservation in this specified service mix.
The capacity feasibility constraint at each day~$d$ is
\begin{equation}
  \sum_{i \in \mathcal{P}(d)} h_i \,\leq\, C,
  \quad h_i \,\geq\, \underline{h}_i.
  \label{eq:cap}
\end{equation}
The grand coalition is feasible iff
$\sum_i \underline{h}_i \leq C$; otherwise no single coalition
honours all SLAs simultaneously, motivating dynamic sub-coalition
formation.

\paragraph{Coalition Value, Veto, and Information}

For feasible coalitions containing the NO, the value is the optimum
of \eqref{eq:char_function}. Coalitions without the NO have value zero.
The analysis and baseline experiments assume
$C>\sum_{i=1}^{M}\underline h_i$, so every SP subset is feasible.
Outside this domain, infeasible sets are excluded from optimization;
an infeasible investor universe is not passed to the Shapley calculation.

\begin{proposition}[Veto player]
\label{prop:veto}
$v_d(\mathcal{S}) = 0$ whenever $\NO \notin \mathcal{S}$.
\end{proposition}
\begin{proof}
The NO controls access to the deployed infrastructure; without
its participation, the coalition cannot operate and has value zero
by the characteristic-function convention.
\end{proof}

Each SP observes its own traffic and revenue coefficient, its
contractual reservation, and a small set of public signals: the NO's value declaration $\hat{v}(d)$, per-SP
marginal contribution estimates $\hat{\Delta}_i(d)$, the active 
coalition size $|\mathcal{S}(d)|$, and its own
$\Lambda^i_d,\beta_i,\underline h_i$. The shadow price $\lambda^*(d)$
of \eqref{eq:char_function} is its capacity multiplier, in vu/mc;
it and other SPs' parameters remain private to the NO.
The declared signals $\hat v(d)$ and $\hat\Delta_i(d)$ estimate
the current coalition value and the entry/departure marginal
$v_d(\mathcal S(d)\cup\{i\})-v_d(\mathcal S(d)\setminus\{i\})$.
Cardinality is denoted by $|\cdot|$.

Evaluating $\Delta_i(d)$ (Definition~\ref{def:Delta}) requires
$v_d$ on sub-coalitions, hence $\lambda^*(d)$ and competitors'
parameters. The NO therefore publishes daily the vector
$\{\hat{\Delta}_i(d)\}_i$ alongside $\hat{v}(d)$, and each SP applies
\eqref{eq:Vleave}--\eqref{eq:MoSjoin} to it without reconstructing
$\lambda^*$. The audit of Section~\ref{sec:audit} covers the full
declared vector $(\hat{v}(d), \{\hat{\Delta}_i(d)\})$: detection on any
\rev[R4.M5]{component triggers the penalty \textcolor{black}{$\Pi$}. The bound in
Section~\ref{sec:sp} covers the induced feasible coalition outcomes
under verified settlement.}

\paragraph{Capacity Regimes and Coalition Structure}
\label{sec:regimes}
For the structural analysis of SP participation, the NO is held
fixed: for any SP set $\mathcal P\subseteq\{1,\ldots,M\}$, we define $w_d(\mathcal P)=v_d(\mathcal P\cup\{\NO\})$.
Submodularity below refers to this set function on SP subsets.
Payoff allocation still uses the Shapley value of the full game,
including the NO as a veto player.
The reference level $q>0$ is a marginal value in vu/mc, fixed at
$0.10$ in the experiments, and distinct from $q_{\mathrm{cap}}$.
The coupling through $C$ induces a regime structure indexed by the
shadow price $\lambda^*(d)\geq0$ of \eqref{eq:char_function}.
In $h_i^*$ and $\lambda^*$, the asterisk marks the optimal allocation and its capacity multiplier. Define $\mathcal P^+(d)=\{i\in\mathcal P(d):h_i^*>\underline h_i\}$ (unconstrained SPs)
and $\mathcal P^0(d)=\{i\in\mathcal P(d):h_i^*=\underline h_i\}$ (SLA-bound SPs).

\begin{definition}[Capacity regimes]
\label{def:regimes}
For the current coalition, write
$\lambda^*(d)=\lambda^*(\mathcal S(d),d)$.
\emph{Regime~A} (abundant): $\lambda^*(d)\leq q$.
\emph{Regime~B} (scarce): $\lambda^*(d)>q$.
These labels describe scarcity relative to the reference level $q$.
Under the assumptions of Theorem~\ref{thm:convex}, the coalition
value conditional on the NO's presence is submodular in both regimes.
Binding SLA floors may produce negative marginal contributions.
For a coalition containing only the NO, we adopt the convention
$\lambda^*(d)=0$ and assign it to Regime~A.
\end{definition}

The comparison between $\lambda^*(d)$ and $q$ describes the
capacity regime. The separability identity in
Proposition~\ref{prop:sep} applies to a specified coalition under
interior allocation; it does not imply coalition-independent marginal
contributions. A fixed membership can be suboptimal when excluding
an SP increases the operational coalition value.


\begin{table*}[t]
\centering\small
\caption{Principal notation and operational parameter set. Per-SP vectors
are listed in the order (Video, AR/VR, AutoDrive, IoT, Gaming). Values match the accompanying reproducibility package.}
\label{tab:notation}
\renewcommand{\arraystretch}{0.92}
\setlength{\tabcolsep}{2.5pt}
\begin{minipage}[t]{0.495\textwidth}
\centering
\begin{tabular}{@{}>{\raggedright\arraybackslash}p{1.5cm} >{\raggedright\arraybackslash}p{3.4cm} >{\raggedright\arraybackslash}p{2.6cm}@{}}
\toprule
\textbf{Symbol} & \textbf{Meaning} & \textbf{Value} \\
\midrule
\multicolumn{3}{@{}l}{\emph{Players and coalitions}} \\
$\mathcal{I}$ & Full player set $\{1,\ldots,M,\NO\}$ & $M = 5$ SPs $+$ 1 NO \\
$\mathcal{P}^+(d)$, $\mathcal{P}^0(d)$ & Unconstrained, SLA-bound SPs & --- \\
$\mathcal{S}^{\mathrm{opt}}(d)$ & Oracle-optimal coalition & --- \\
\midrule
\multicolumn{3}{@{}l}{\emph{Time}} \\
$D_{\mathrm{inv}}$ & Investment horizon [day] & $365$, $1095$ \\
$T$ & Intra-day timeslots & $96$ \\
\midrule
\multicolumn{3}{@{}l}{\emph{Traffic process}} \\
$\Lambda^i_0$ & Initial traffic scale [req/s] & $(400, 250, 150,$ $800, 350)$ \\
$\Lambda^i_d$ & Daily traffic scale [req/s] & GRW, Eq.~\eqref{eq:grw} \\
$\mu_i$ & GRW drift [1/day] & $(8, 12, 5, 6, 10)$ $\times 10^{-4}$ \\
$\sigma_i$ & GRW volatility [$1/\sqrt{\text{day}}$] & $(40, 60, 50,$ $30, 45) \times 10^{-3}$ \\
$F$ & Intra-day profile sum & $96.0$ \\
\midrule
\multicolumn{3}{@{}l}{\emph{Capacity and SLA}} \\
$C$ & Deployed capacity [mc, millicores] & $3100.394$ \\
$q$ & Reference marginal value [vu/mc] & $0.10$ \\
$\xi$ & Saturation parameter [1/mc] & $10^{-3}$ \\
$\underline{h}_i$ & SLA floor [mc] & $(30, 300, 1500,$ $0.3, 112.5)$ \\
$\sum_i \underline{h}_i$ & Aggregate SLA floor [mc] & $1942.8$ ($62.7\%$ of $C$) \\
$h^*_i(d)$ & Optimal allocation [mc] & Eq.~\eqref{eq:hstar} \\
$\lambda^*(d)$ & Active shadow price [vu/mc] & Coalition-dependent \\
\midrule
\multicolumn{3}{@{}l}{\emph{Utility and revenue}} \\
$U^i_d(h_i)$ & Daily utility [vu] & Eq.~\eqref{eq:Ud} \\
$\beta_i$ & Normalized value coefficient [vu/(req/s)] & $(8, 15, 25,$ $4, 12) \times 10^{-3}$ \\
\bottomrule
\end{tabular}
\end{minipage}%
\hfill
\begin{minipage}[t]{0.495\textwidth}
\centering
\begin{tabular}{@{}>{\raggedright\arraybackslash}p{1.5cm} >{\raggedright\arraybackslash}p{3.4cm} >{\raggedright\arraybackslash}p{2.6cm}@{}}
\toprule
\textbf{Symbol} & \textbf{Meaning} & \textbf{Value} \\
\midrule
\multicolumn{3}{@{}l}{\emph{Coalition value}} \\
$v_d(\mathcal{S})$ & Constrained coalition value [vu] & Eq.~\eqref{eq:char_function} \\
$\tilde{v}_d(\mathcal{S})$ & Net value (cost-adjusted) [vu] & Eq.~\eqref{eq:vnet} \\
$v^i(d)$ & Unconstrained individual value & Eq.~\eqref{eq:vi} \\
$A_i,\,X_i(d)$ & Coefficient, profitability index & Eq.~\eqref{eq:Xi} \\
\midrule
\multicolumn{3}{@{}l}{\emph{Shapley value}} \\
$\Delta_i(d)$ & Marginal contribution [vu] & Def.~\ref{def:Delta} \\
$\varphi^+_i(d)$ & Benchmark Shapley value & Eq.~\eqref{eq:shapley_A} \\
$\varphi^-_i(d)$ & Operational Shapley value & Eq.~\eqref{eq:shapley_B} \\
$R,\,K$ & Hoeffding bound, sample size & Eq.~\eqref{eq:hoeffding} \\
\midrule
\multicolumn{3}{@{}l}{\emph{Value-investing rule}} \\
$V_i(d)$ & Value ratio (active SP) & Eq.~\eqref{eq:Vleave} \\
$\MoS_i(d)$ & Margin of safety (inactive SP) & Eq.~\eqref{eq:MoSjoin} \\
$\theta_{\min},\,\theta^-_{\min}$ & Base margins & $0.050$, $0.025$ \\
$\kappa,\,\kappa^-$ & Risk-aversion multipliers & $0.10$, $0.05$ \\
$\zeta^{\mathrm{join}}_i$, $\zeta^{\mathrm{leave}}_i$ & Observed action times & --- \\
\midrule
\multicolumn{3}{@{}l}{\emph{Audit and trust}} \\
$\alpha$ & Nominal audit frequency & $0.15$ \\
$\alpha^*$ & Fixed-state criterion optimum & $0.9140$ \\
$B_d$ & Operational gain upper bound [vu] & Mean $114.201$ \\
$\Pi^{\mathrm{suff}}_d$ & Sufficient penalty [vu] & Mean $10.745$ at $\alpha^*$ \\
$c_a$ & Audit cost per occurrence [vu] & $22.5$ \\
$\rho$ & Trust risk aversion & $0.05$ \\
$\hat{\delta}(d),\,\tau(d)$ & Audit pass rate, trust factor & Eq.~\eqref{eq:trust} \\
$\varepsilon$ & Scalar test perturbation & $-15\%$ \\
\midrule
\multicolumn{3}{@{}l}{\emph{Efficiency metrics and reproducibility}} \\
$\PoD,\,\PoS$ & Decentr. price, dynamic/static ratio & Eq.~\eqref{eq:podpos} \\
--- & Random seed & $4$ \\
\bottomrule
\end{tabular}
\end{minipage}

\par\smallskip
\begin{minipage}{\textwidth}
\footnotesize
\textit{Abbreviations:} NO, Network Operator; SP, Service Provider;
SLA, service-level agreement; GRW, Geometric Random Walk;
TU, transferable utility; MoS, margin of safety;
PoD, Price of Decentralisation; PoS, dynamic/static value ratio;
KKT, Karush--Kuhn--Tucker; CV, coefficient of variation;
FOC, first-order condition;
CAPEX, capital expenditure; AR/VR, augmented/virtual reality;
IoT, Internet of Things;
mc, millicore; vu, normalized value unit.
\end{minipage}
\end{table*}
\section{Analysis}
\label{sec:analysis}

\paragraph{Karush--Kuhn--Tucker (KKT) Solution}

Let $q>0$ be a reference marginal value in vu/mc, fixed at $0.10$;
it is distinct from the hardware cost $q_{\mathrm{cap}}$.
Define the coefficient $A_i$ (units s/req) and the dimensionless
profitability index $X_i(d)$ by
\begin{equation}
  A_i = \frac{\xi\,\beta_i\,F}{q}, \qquad
  X_i(d) = A_i\,\Lambda^i_d.
  \label{eq:Xi}
\end{equation}
$X_i=1$ is the break-even point for the first infinitesimal
millicore in the unconstrained reference-cost benchmark; it is not
a test of profitability for a mandatory SLA reservation.

\begin{proposition}[KKT solution]
\label{prop:kkt}
For a coalition containing the NO and at least one SP, with
$C>\sum_{i\in\mathcal P(\mathcal S)}\underline h_i$, the optimum
of \eqref{eq:char_function} is
\begin{equation}
  h^*_i(d) = \max\!\left\{\underline{h}_i,\;
    \frac{1}{\xi}\ln\frac{X_i(d)\,q}{\lambda^*(d)}\right\},
  \label{eq:hstar}
\end{equation}
with $\lambda^*(d) > 0$ uniquely determined by
$\sum_i h^*_i(d) = C$.
\end{proposition}
The proof is given in Appendix~\ref{app:kkt}.

The interior solution admits a water-filling reading: the allocator
pours capacity into each SP until its marginal utility, equal to $X_iq e^{-\xi h_i^*}$, equals the common shadow price
for every SP allocated strictly above its floor.

\paragraph{Unconstrained Individual Value and Separability}

For $x>0$, define $g(x)=x-1-\ln x$ ($g\geq0$, $g(1)=0$).
The \emph{unconstrained individual value} of SP~$i$ is
\begin{equation}
{
v^i(d)=\begin{cases}
\dfrac{q}{\xi}g(X_i(d)),&X_i(d)>1,\\
0,&0<X_i(d)\leq1.
\end{cases}}
\label{eq:vi}
\end{equation}
which equals the surplus that SP~$i$ would extract if it could buy
capacity at the reference value $q$ alone (derivation in Appendix~\ref{app:kkt}). The
auxiliary cost-adjusted value is

\begin{equation}
  \tilde{v}_d(\mathcal{S}) = v_d(\mathcal{S})
  - q \sum_{i \in \mathcal{P}(\mathcal{S})} h^*_i(\mathcal{S}, d).
  \label{eq:vnet}
\end{equation}

Stage~II redistributes the operational value $v_d$.
The cost-adjusted quantity $\tilde v_d$ is used only
for comparison with the individual-value benchmark;
the deployment CAPEX is accounted for in Stage~I.

\begin{proposition}[Coalition-specific separability identity]
\label{prop:sep}
Fix a feasible coalition $\mathcal S$ containing the NO and at
least one SP. Assume that all its SP allocations are strictly above
their SLA floors and that $X_i(d)>1$ for every member. If
$\lambda^*(\mathcal S,d)=q$,
\begin{equation}
{\tilde v_d(\mathcal S)=
\sum_{i\in\mathcal P(\mathcal S)}v^i(d).}
\label{eq:sep_exact}
\end{equation}
Under the same assumptions, the discrepancy
is given by \eqref{eq:sep_offedge} and is quadratic as
$\lambda^*(\mathcal S,d)\to q$. This identity concerns the specified
coalition; its sub-coalitions generally have different shadow prices.
\end{proposition}
The proof is given in Appendix~\ref{app:sep}.

\paragraph{Conditional Submodularity}

\begin{theorem}[Conditional submodularity]
\label{thm:convex}
{
Fix day $d$ and assume $C>\sum_{i=1}^{M}\underline h_i$.
For the utilities in \eqref{eq:Ud}, the conditional coalition value
$w_d$ is submodular: for $\mathcal A\subseteq\mathcal B$
and $j\notin\mathcal B$,
\[
w_d(\mathcal A\cup\{j\})-w_d(\mathcal A)
\geq w_d(\mathcal B\cup\{j\})-w_d(\mathcal B).
\]
This property holds independently of the comparison between
$\lambda^*$ and $q$. Positive SLA floors may make $w_d$
non-monotone, allowing negative marginal contributions.
}
\end{theorem}
The proof of the above theorem is in Appendix~\ref{app:convex}.

\section{Dynamic Shapley Value}
\label{sec:shapley}
\begin{theorem}[Benchmark and operational Shapley values]
\label{thm:shapley}
Let $m=|\mathcal S(d)|\geq2$ be the number of active players, including the NO.
\textnormal{(i)} \emph{Separable benchmark.}
For a sub-coalition $\mathcal T\subseteq\mathcal S(d)$, let
$\bar v_d(\mathcal T)$ denote the following auxiliary veto-game value
{
\[
\bar v_d(\mathcal T)=\begin{cases}
\displaystyle\sum_{i\in\mathcal T\setminus\{\NO\}}v^i(d),
&\NO\in\mathcal T,\\
0,&\NO\notin\mathcal T.
\end{cases}
\]}
This game is supermodular and has a nonempty core.
Its Shapley allocation is:
\begin{equation}
\varphi_i^+(d)=\frac{v^i(d)}{2},\quad
\varphi_{\NO}^+(d)=\frac12\sum_{i\in\mathcal P(\mathcal S(d))}v^i(d).
\label{eq:shapley_A}
\end{equation}
\textnormal{(ii)} \emph{Operational game.}
For $v_d$ in \eqref{eq:char_function}, in either capacity regime,
let $\Omega_m$ be the set of player orderings, $\pi\in\Omega_m$
one ordering, and $\mathcal S_i^\pi$ the players preceding $i$.
For $i\notin\mathcal T$, write
$\Delta_i(\mathcal T,d)=v_d(\mathcal T\cup\{i\})-v_d(\mathcal T)$.
The Shapley value is
\begin{equation}
\varphi_i^-(d)=\frac1{m!}\sum_{\pi\in\Omega_m}
\Delta_i(\mathcal S_i^\pi,d).
\label{eq:shapley_B}
\end{equation}
The superscripts $+$ and $-$ identify the benchmark and operational
games, respectively; they do not indicate payoff signs.
For the realized coalition, $\varphi_i(d)=\varphi_i^-(d)$ in
the Stage-I contract. The symbol $\varphi_{\NO}^{\pm}$ denotes
the NO's allocation in the corresponding game. A player's Shapley value is negative
precisely when its permutation-averaged marginal contribution is negative.
\end{theorem}
The proof is given in Appendix~\ref{app:shapley}.
\paragraph{Scalability Beyond the Enumerated Instance}
Caching all coalition values requires $O(2^M)$ allocation
solves; evaluating every Shapley component by weighted subsets
requires $O(M2^M)$ arithmetic operations for one active game and is used
here since $M = 5$. Computing the full payoff table for
all active SP subsets, as required by the audit bound, takes
$O(M3^M)$ weighted-subset operations. For larger $M$, the exact sum is replaced by the
Monte~Carlo estimator~\cite{castro2009shapley}, optionally
stratified~\cite{mitchell2022sampling}, avoiding exhaustive enumeration for a fixed sample count.
\rev[MR.1]{A sampled payoff near zero requires an uncertainty check
before its sign can be interpreted (Remark~\ref{rem:access_fee}).}
\begin{proposition}[Sampling complexity]
\label{prop:shapley_scale}
Let $R=\sum_{j=1}^{M}\beta_j\Lambda_d^jF$. Since each
operational coalition value lies in $[0,R]$, every marginal
contribution lies in $[-R,R]$. Here $R$ bounds the magnitude of a marginal contribution;
$K$ is the sample count, $\delta_{\mathrm{err}}>0$ the additive-error tolerance,
and $\eta_{\mathrm{fail}}\in(0,1)$ the sampling failure probability.
The notation $\Pr$ denotes probability.
Drawing $K$ i.i.d.\ permutations
$\pi_1,\dots,\pi_K$ uniformly and averaging the marginal
contributions of SP~$i$ yields an unbiased estimator
$\hat{\varphi}^-_i(d)$ with, by Hoeffding's inequality,
\begin{equation}
  \Pr\!\bigl(|\hat{\varphi}^-_i(d) - \varphi^-_i(d)| \geq \delta_{\mathrm{err}}\bigr)
  \;\leq\; 2\exp\!\Bigl(-\tfrac{K\delta_{\mathrm{err}}^2}{2R^2}\Bigr),
  \label{eq:hoeffding}
\end{equation}
so $K = O(R^2 \delta_{\mathrm{err}}^{-2}\log(1/\eta_{\mathrm{fail}}))$ permutations
suffice for additive accuracy $\delta_{\mathrm{err}}$ with probability
$1-\eta_{\mathrm{fail}}$, giving total cost $O(K \cdot M)$ water-filling
solves, independent of $2^M$.
\end{proposition}
\begin{proof}
Each sampled marginal contribution lies in $[-R, R]$ and the
permutation is uniform, so the sample mean is unbiased for
\eqref{eq:shapley_B}; \eqref{eq:hoeffding} is Hoeffding's bound for
bounded i.i.d.\ terms. Inverting the exponential gives the stated
$K$. For simultaneous accuracy for all $M+1$ players, replace the
failure probability by $\eta_{\mathrm{fail}}/(M+1)$ using the union bound.
The implementation reports this simultaneous Hoeffding radius.
\end{proof}
\begin{remark}[{\rev[MR.1]{Interpretation of negative Shapley payoffs}}]
\label{rem:access_fee}
\rev[MR.1]{A negative operational Shapley value is a theoretical
payoff-allocation signal: the SP's marginal contribution, averaged over
player orderings, is negative. In the TU model,
$\varphi^-_i(d)<0$ prescribes an outgoing transfer of magnitude
$|\varphi^-_i(d)|$; it does not by itself imply a monetary charge.}
The participation rule responds to marginal-contribution signals;
it does not explicitly optimize cumulative fees or future individual
payoffs. Negative Shapley and the departure criterion need not coincide.
\rev[MR.1]{Implementing this transfer as an access fee,
resource-reservation charge or SLA-related cost requires a contract
specifying payment, enforcement and recovery terms. Truthful declarations
remain subject to the audit assumptions of Theorem~\ref{thm:sp}.}
\end{remark}
\section{Properties of the Shapley Value Process}
\label{sec:shapley_props}
The benchmark Shapley payoff is a nonlinear function of
log-normal traffic. In the high-demand limit, $g(X_i)\sim X_i$,
giving an approximately log-normal payoff.
The following calculation concerns terminal traffic uncertainty,
not the variance of cumulative operational Shapley payments. The key result, used in Section~\ref{sec:vi} to
calibrate the margin-of-safety threshold, is the residual-horizon
volatility.
Intuitively, an SP's benchmark payoff is approximately proportional
to its own traffic at high demand, so its uncertainty compounds with the volatility
$\sigma_i$ and the length of the residual horizon; the margin of
safety must therefore widen with both.
\begin{proposition}[Coefficient of variation (CV)]
\label{prop:CV}
Let $D_{\mathrm{res}}=D_{\mathrm{inv}}-d$ be the residual horizon.
For a positive random variable $Z$, define
$\CV(Z)=\sqrt{\Var(Z)}/\mathbb E[Z]$, where $\Var$ and
$\mathbb E$ denote variance and expectation.
Conditional on the current traffic scale, the GRW gives
\begin{equation}
\CV(\Lambda^i_{d+D_{\mathrm{res}}}\mid\Lambda^i_d)
=\sqrt{e^{\sigma_i^2D_{\mathrm{res}}}-1}
\approx\sigma_i\sqrt{D_{\mathrm{res}}}.
\label{eq:CV}
\end{equation}
The last approximation requires $\sigma_i^2D_{\mathrm{res}}\ll1$.
It motivates a benchmark-payoff approximation only when the
high-demand approximation $g(X_i)\approx X_i$ is also suitable.
\end{proposition}
\begin{proof}
The conditional log increment has variance
$\sigma_i^2D_{\mathrm{res}}$. The log-normal moment formula gives
the exact CV in \eqref{eq:CV}. Expanding $e^x-1$ at zero gives
the stated small-variance approximation.
\end{proof}
\section{Value-Investing Decision Rule}
\label{sec:vi}
\paragraph{Value Ratio and Margin of Safety}
\begin{definition}[Marginal contribution]
\label{def:Delta}
$\Delta_i(d) = v_d(\mathcal{S}(d) \cup \{i\})
              - v_d(\mathcal{S}(d) \setminus \{i\})$.
For $i \in \mathcal{S}(d)$ this is the value lost on departure;
for $i \notin \mathcal{S}(d)$, the value gained on entry.
\end{definition}
\begin{definition}[Value ratio and margin of safety]
\label{def:Vratio}
The published estimates satisfy $\hat v(d)\geq0$, and
$\hat\Delta_i(d)$ estimates the entry/departure marginal.
The dimensionless trust multiplier $\tau(d)$ is defined in
Section~\ref{sec:audit}. For $\hat v(d)>0$, define the value ratio
$V_i(d)$ for active $i\in\mathcal P(d)$ by
\begin{equation}
  V_i(d) = \frac{|\mathcal{P}(d)|\,\hat\Delta_i(d)\,\tau(d)}{\hat{v}(d)}.
  \label{eq:Vleave}
\end{equation}
For an inactive SP, define the margin-of-safety ratio $\MoS_i(d)$ by
\begin{equation}
  \MoS_i(d) = \frac{\hat\Delta_i(d)}{\hat{v}(d)}\,\tau(d),
  \label{eq:MoSjoin}
\end{equation}
with $\hat{v}(d)$ the NO's declared coalition value and $\tau(d)$ the
trust factor of Section~\ref{sec:audit}. Both rules use the
declared $\hat{v}(d)$ rather than the unobservable true value
$v_d(\mathcal{S}(d))$, consistent with the partial-information
structure of Section~\ref{sec:model}. In \eqref{eq:Vleave} the ratio
$|\mathcal{P}(d)|\,\hat\Delta_i(d)/\hat{v}(d)$ measures SP~$i$'s
contribution relative to the average member contribution
$\hat{v}(d)/|\mathcal{P}(d)|$.
\end{definition}
The Graham analogy motivates the choice of marginal contribution
as a value signal. The leave rule compares it with average current
coalition value, whereas the join rule compares the incremental value
with a fraction of total coalition value. These are distinct
operational criteria, not a derived maximization of individual
Shapley payoff.
\rev[R3.M1b]{Average coalition value supplies a common scale for the
marginal signal, not a fair-share entitlement. This choice can remove
positive contributors; Section~\ref{sec:results} reports the resulting
loss when capacity is abundant.}
\begin{definition}[Join and leave rules]
\label{def:joinleave}
At day $d$, $\theta_i(d)>0$ and $\theta_i^-(d)>0$ denote
the join and leave margin parameters, specified below.
\emph{join} ($i \notin \mathcal{S}(d)$):
$\MoS_i(d) \geq \theta_i(d)$, $\hat\Delta_i(d)>0$, and
$\sum_{j \in \mathcal{P}(\mathcal{S}(d)) \cup \{i\}} \underline{h}_j \leq C$;
\emph{leave} ($i \in \mathcal{S}(d)$):
$V_i(d)\leq1/(1+\theta_i^-(d))$.
When $\hat v(d)=0$, the rules are evaluated by cross
multiplication: join requires $\hat\Delta_i(d)>0$ and
$\tau(d)\hat\Delta_i(d)\geq\theta_i(d)\hat v(d)$; leave requires
$|\mathcal P(d)|\tau(d)\hat\Delta_i(d)\leq
\hat v(d)/(1+\theta_i^-(d))$.
\rev[R3.M1c]{Each operational day starts from the preceding day's coalition.
The implementation processes exits, then entries, in the fixed order
Video, AR/VR, AutoDrive, IoT, Gaming, recomputing coalition values
after each change. An SP may exit and re-enter within the same day.
Only this single ordered sweep is used for the reported daily values;
the frozen-day iterations below are diagnostic. Baseline declarations
are truthful, \textcolor{black}{$\hat v=v_d(\mathcal S)$} and \textcolor{black}{$\hat\Delta_i=\Delta_i$}. Exits precede entries so that released
capacity is available to prospective entrants.}
\end{definition}
\paragraph{Threshold Calibration}
\begin{definition}[Threshold calibration]
\label{prop:theta}
The join threshold is
\begin{equation}
\theta_i(d)=\theta_{\min}+\kappa\sigma_i\sqrt{D_{\mathrm{inv}}-d},
\label{eq:theta}
\end{equation}
where $\theta_{\min}>0$ is the base margin and $\kappa>0$ is a
dimensionless sensitivity parameter.
This is a design rule motivated by the small-variance scaling
in Proposition~\ref{prop:CV}; it is not an exact operational-payoff
risk bound. The parameters remain tunable, and their sensitivity
is evaluated numerically.
\end{definition}
\begin{remark}[Asymmetric thresholds]
\label{rem:asym_theta}
The leave threshold uses $\kappa^- < \kappa$:
$\theta^-_i(d) = \theta^-_{\min} + \kappa^- \sigma_i \sqrt{D_{\mathrm{inv}} - d}$,
where $\theta^-_{\min}>0$ is the leave base margin and
$\kappa^->0$ its volatility coefficient; the experiments use
$\theta^-_{\min}<\theta_{\min}$ and $\kappa^-<\kappa$.
These coefficients set different admission and retention
criteria. Since the two rules use different ratios, their asymmetry
does not by itself prove a hysteresis band or prevent within-day
exit and re-entry. Experiments
use $\kappa = 0.10$, $\kappa^- = 0.05$.
\end{remark}
\section{Probabilistic Auditing Mechanism}
\label{sec:audit}
\paragraph{Trust Factor}
Let $\alpha\in(0,1]$ be the audit probability,
$\Pi\geq0$ the per-instance penalty, $c_a>0$ the cost of one audit,
and $\rho\in(0,1)$ the trust parameter. Penalties and costs use vu.
At each day, an independent audit is conducted with probability $\alpha$. Conditional on auditing, any misreport is
detected with probability~1; on detection the inflated outcome is
voided and the NO pays a per-instance penalty $\Pi$~[vu].
Let $n^{\mathrm{pass}}(d)$ and $n^{\mathrm{audit}}(d)$ count passed
and completed audits available before day $d$'s decisions.
Each SP maintains the Laplace-smoothed audit pass rate
\(  \hat{\delta}(d) = \frac{n^{\mathrm{pass}}(d) + 1}
                         {n^{\mathrm{audit}}(d) + 2},\)
where $\mathrm{Beta}(1,1)$ is the uniform prior for the audit-pass
probability. The smoothed rate is its posterior mean. Smoothing
gives $\hat{\delta}(0) = 1/2$ for SPs entering before any audit,
preventing collapse on a single early failure. The trust factor is
\begin{equation}
  \tau(d) = 1 - \rho^{\hat{\delta}(d)} \in [0, 1-\rho],
  \label{eq:trust}
\end{equation}
with $\rho \in (0,1)$ a risk-aversion parameter; $\tau$ increases
strictly in $\hat\delta$. With no audit history,
$\tau=1-\sqrt{\rho}$; with an increasing all-pass history,
$\tau$ approaches $1-\rho$ from below. The baseline uses $\rho=0.05$.
An audit performed after day $d$'s decisions updates trust for
day $d+1$. Delayed or erroneous trust updates are studied separately
from the idealized one-day financial settlement.
\paragraph{Numerator Placement of \texorpdfstring{$\tau$}{tau}}
Discounting the declaration instead ($\hat{v} \to \tau\hat{v}$) would
be perverse for positive marginal signals: as $\tau \to 0$ the denominator of $V_i$ vanishes, $V_i$
diverges, and joins are \emph{encouraged}. For positive marginal signals, placing $\tau$ in the numerator
reduces admission and retention ratios when trust falls. This can
change coalition membership, but it does not by itself prove that
the NO loses value, since an excluded SP may have a negative marginal
contribution.
\section{Equilibrium Analysis}
\label{sec:equil}
\begin{definition}[Observed stopping times]
\label{thm:stopping}
For action $a\in\{\mathrm{join},\mathrm{leave}\}$, let
$\zeta_i^a$ be the first operational day on which the ordered rule
executes action $a$ for SP~$i$. If no such action occurs by
$D_{\mathrm{inv}}$, set $\zeta_i^a=\infty$ and record it as
unobserved within the horizon. No almost-sure eventual action is
asserted.
\end{definition}
\begin{definition}[Value-investing equilibrium]
\label{def:VIeq}
An active coalition $\mathcal{S}^*(d)$ is an equilibrium if no active SP wishes to
leave and no inactive SP wishes to join (no feasible join or leave action in
Definition~\ref{def:joinleave} is eligible).
\end{definition}
\begin{theorem}[Finite detection of repeated states]
\label{thm:exist}
Fix day $d$, its traffic, trust and thresholds. Let
$\mathsf B$ denote one deterministic ordered sweep and $k$ the
sweep index. Iterating
$\mathcal P_{k+1}=\mathsf B(\mathcal P_k)$ on the power set
$2^{\{1,\ldots,M\}}$ reaches a repeated state after at most $2^M$
transitions. The resulting orbit has some period $p\geq1$.
A period-one sweep state need not satisfy
Definition~\ref{def:VIeq}, because exits and entries can cancel
within a sweep.
\end{theorem}
\begin{proof}
Among $2^M+1$ successive states, two coincide. Determinism
then makes the subsequent orbit periodic. Equality of sweep
endpoints does not exclude intermediate membership changes.
\end{proof}
\begin{proposition}[Conditional convergence to equilibrium]
\label{prop:tarski}
\rev[R3.M1a,R3.m2,R2.3.2]{Fix the day's traffic, trust, thresholds,
SP order and truthful signals. Let \textcolor{black}{$\mathcal P_0$} be the initial feasible SP set and
\textcolor{black}{$\mathcal K$} the sets reachable from \textcolor{black}{$\mathcal P_0$} through eligible individual moves. For each SP~\textcolor{black}{$i$}, let
\textcolor{black}{$L_i$} be its largest departure marginal over eligible exits in
\textcolor{black}{$\mathcal K$}, and \textcolor{black}{$J_i$} its smallest entry marginal over eligible
entries. Use \textcolor{black}{$L_i=-\infty$} or \textcolor{black}{$J_i=+\infty$} for an empty exit or
entry set. If \textcolor{black}{$L_i<J_i$} for every SP, repeated sequential sweeps
from \textcolor{black}{$\mathcal P_0$} reach an equilibrium after at most \textcolor{black}{$|\mathcal K|-1$} membership changes.}
\end{proposition}
\begin{proof}
Choose finite constants $a_i$ with $L_i<a_i<J_i$ and define the
potential $\Phi(\mathcal P)=w_d(\mathcal P)-\sum_{i\in\mathcal P}a_i$.
Each eligible entry increases it by $\Delta_i-a_i>0$; each eligible
exit by $a_i-\Delta_i>0$. Thus no set repeats after an eligible move.
Every sweep tests all SPs, so termination implies no eligible move.
\end{proof}
For seed 4 over 1095 days, frozen-day diagnostics find 1093
period-one orbits and two period-two orbits, with mean 1.128 and
maximum three transitions to first repetition. Only 97 of the 1095
realized daily endpoints satisfy Definition~\ref{def:VIeq}. Exact enumeration finds no individual equilibrium on the other 998
days. The same 97 days satisfy the potential condition on reachable
sets. A general convergence characterization remains future work.
\section{Strategy-Proofness}
\label{sec:sp}
\begin{lemma}[NO's gain from benchmark expansion]
\label{lem:incentive}
In the separable benchmark, if a declaration induces a feasible
expansion by a set $\mathcal J$ of SPs without departures, the NO's
gain is $\frac12\sum_{i\in\mathcal J}v^i(d)$, strictly positive
when this sum is positive. This statement does not exclude profitable
misreporting in either regime of the operational game.
\end{lemma}
\begin{proof}
The result follows by subtracting the NO's benchmark allocation
before entry from its allocation after entry in \eqref{eq:shapley_A}.
\end{proof}
\begin{theorem}[Strategy-proofness]
\label{thm:sp}
Fix the current state and other players' responses. For a
risk-neutral NO maximizing its current-day expected payoff, assume
perfect detection upon audit, no penalty for truthful reporting,
and rollback to the truthful outcome minus $\Pi$. Let $\mathcal R_d$
be the set of admissible false declarations and $\mathbf r$ a complete
declaration of coalition values and marginal signals within the sweep. Truth-telling is weakly optimal against every admissible
misreport iff
\begin{equation}
  \Pi \;\geq\; \frac{1 - \alpha}{\alpha}\,
   \sup_{\mathbf r\in\mathcal R_d} G(\mathbf r),
  \label{eq:SP_condition}
\end{equation}
Here $G(\mathbf r)$ is the NO's operational Shapley gain if
declaration $\mathbf r$ goes undetected. The scalar $\varepsilon$ is
reserved for the experiment $\hat v=(1+\varepsilon)v$, with truthful
marginal signals. For a benchmark expansion without departures,
$G^+(\mathbf r)=\frac12\sum_{i\in\mathcal J(\mathbf r)}v^i(d)$.
For the operational game, $G$ uses the induced coalition and its
verified operational Shapley allocation. Reports affect admission
signals, not the settlement formula itself. This excludes direct
payment diversion and assumes enforceable settlement.
\end{theorem}
The proof is given in Appendix~\ref{app:audit}.
\begin{corollary}[Minimum penalty and optimal audit frequency]
\label{cor:opt_audit}
Define the nonnegative worst-case operational daily gain by
\begin{equation}
{\Pi_0^*(d)=\max\{0,\sup_{\mathbf r\in\mathcal R_d}G(\mathbf r;d)\}.}
\label{eq:Pi0star}
\end{equation}
In the separable benchmark, this equals
$\frac12\sum_{i\notin\mathcal S(d)}v^i(d)$ only if the admissible
reports induce expansions without departures and can jointly admit
all positive-value inactive SPs.
The corresponding strategy-proof penalty threshold from
Theorem~\ref{thm:sp} is then
\begin{equation}
  \Pi^*(\alpha;d)=\frac{1-\alpha}{\alpha}\,\Pi_0^*(d).
\label{eq:Pistar}
\end{equation}
Letting $c_a>0$~[vu] be the cost of one audit and assuming $0<\mathbb E[\Pi_0^*]<\infty$ for a fixed distribution
of reference states, the audit frequency minimizing the specified design criterion
$C_{\mathrm{audit}}(\alpha) = \alpha\,c_a +
 (1 - \alpha)\,\mathbb{E}[\Pi^*(\alpha; d)]$
over $\alpha \in (0, 1]$ is
\begin{equation}
  \alpha^* \;=\;
   \sqrt{\frac{\mathbb{E}[\Pi_0^*]}{c_a + \mathbb{E}[\Pi_0^*]}}.
  \label{eq:alphastar}
\end{equation}
For implementation, let $B_d$ be the maximum increase in the NO's
verified operational payoff over all feasible destination coalitions,
relative to the truthful outcome. For the Stage-I contract, use
total receipts $x_{\NO}$ instead of the operational account when
bounding deviations; the numerical audit retains the operational account.
This gives $B_d\geq\Pi_0^*(d)$,
possibly strictly, because some destinations may be unreachable.
Replacing $\Pi_0^*(d)$ by $B_d$ in \eqref{eq:Pistar} gives a sufficient
daily penalty. For conditional detection probability $p_{\mathrm{det}}\in(0,1]$,
replace $\alpha$ by $\alpha p_{\mathrm{det}}$ under the same immediate
rollback assumptions. Numerical calibration is reported in
Section~\ref{sec:results}; the bound is not asserted to be the minimum
over attainable reports. Full derivation in Appendix~\ref{app:audit}.
\end{corollary}
\section{Price of Decentralisation}
\label{sec:pod}
\begin{definition}[Oracle-optimal coalition]
\label{def:Sopt}
$\mathcal{S}^{\mathrm{opt}}(d) =
\argmax_{\mathcal{S} \text{ feasible}} v_d(\mathcal{S})$,
the same-day full-information centralised optimum.
$\mathcal{S}^{\mathrm{opt}}(d)$ knows $\lambda^*(d)$ and
$\{\Lambda^i_d\}$ but is \emph{not} clairvoyant about future demand.
Computing it requires every SP's private parameters, unavailable to
individual SPs.
\end{definition}
\begin{definition}[Price of Decentralisation and dynamic/static ratio]
The label $\PoS$ denotes this paper's dynamic/static value ratio;
it is not the standard equilibrium-based Price of Stability.
\label{def:podpos}
\begin{equation}
  \PoD = \frac{\sum_d v_d(\mathcal{S}^{\mathrm{opt}}(d))}
              {\sum_d v_d(\mathcal{S}(d))} \geq 1,
  \quad
  \PoS = \frac{\sum_d v_d(\mathcal{S}(d))}{\sum_d v_d(\mathcal{I})}.
  \label{eq:podpos}
\end{equation}
$\PoD$ close to 1 means the rule recovers most of the achievable
value despite partial information. $\PoS > 1$ means the dynamic
coalition strictly outperforms the static grand coalition, an outcome enabled by non-monotonicity of the operational
coalition value under mandatory SLA allocations.
\end{definition}
\section[Numerical Results]{Numerical Results\protect\footnote{Code and data:
\url{https://github.com/patan3saro/EdgeValueInvesting-Sim}.}}
\label{sec:results}
\begin{figure}[t]
\centering
\includegraphics[width=\columnwidth]{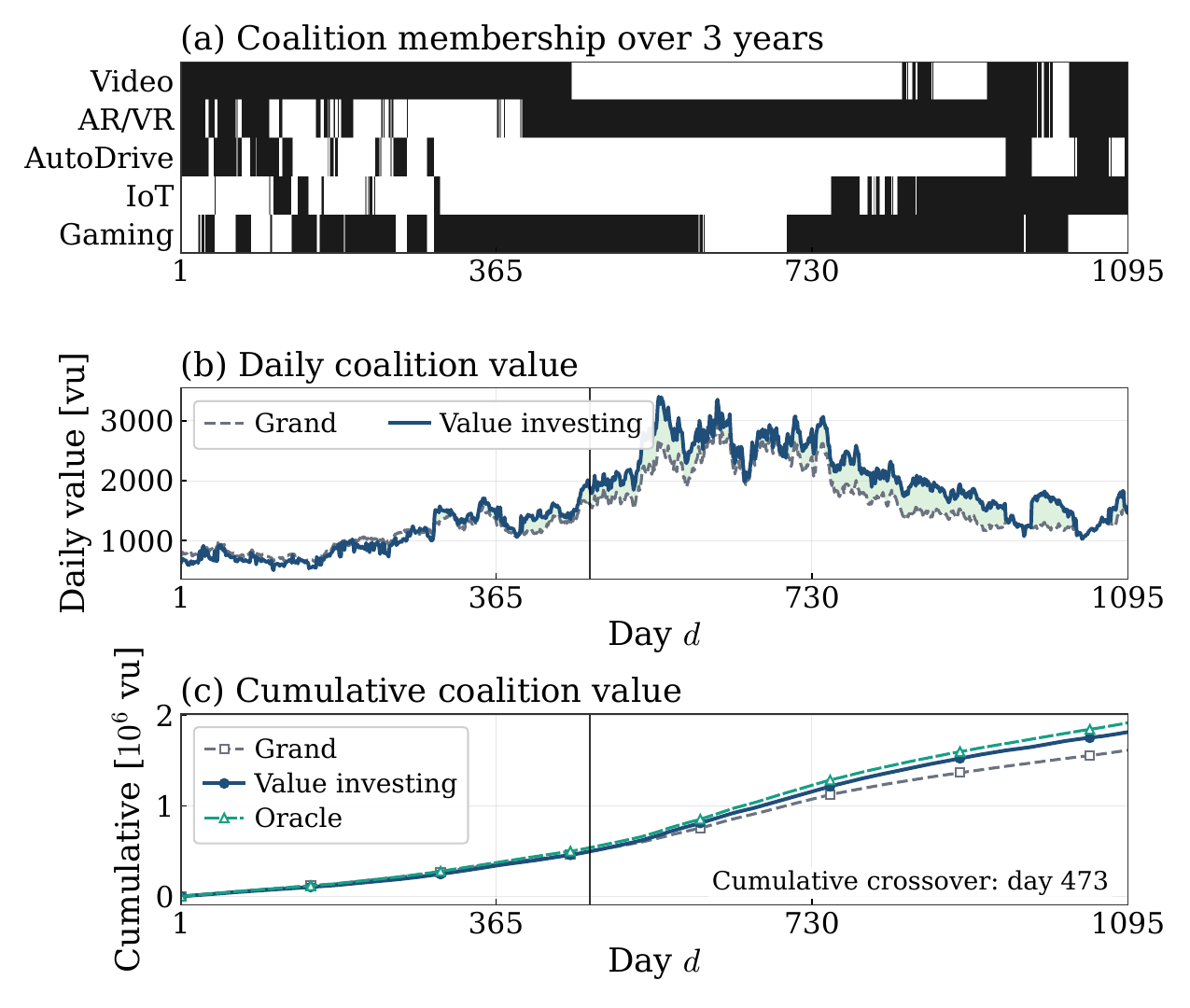}
\caption{Three-year horizon: membership, daily value, and
cumulative comparison. The dynamic coalition overtakes the static
grand coalition at $d=473$. Final $\PoS = 1.124$,
$\PoD = 1.057$.}
\label{fig:3yr}
\end{figure}

\paragraph{Setup and Operating Regime}
The simulations use $M=5$ SPs corresponding to video streaming,
AR/VR, autonomous driving, IoT telemetry and cloud gaming.
Capacity is $C=3100.394$ mc (rounded), the aggregate reservation
is $1942.8$ mc, and horizons are $D_{\mathrm{inv}}\in\{365,1095\}$ days.
The allocator uses the sorted active-set solution of the KKT system;
tests compare its objective and feasibility with an independent
constrained numerical optimizer. All daily coalition values and
operational Shapley allocations are evaluated exactly.
The complete parameter set is in Table~\ref{tab:notation}.
Results aggregate operational days $1,\ldots,D_{\mathrm{inv}}$;
day zero is initialization. Thirty seeds, $0,\ldots,29$, are used,
with seed 4 retained for the detailed trajectory. Strategies share
the same traffic path within each seed and horizon. The two horizons
use the original horizon-dependent random-array protocol and are
not common prefixes of one trajectory. Grand-coalition shadow prices
exceed $q$ on all 1095 days for seed 4, with mean $0.4905$ vu/mc;
active-coalition prices exceed $q$ on $96.895\%$ of days.
\paragraph{Three-Year Run}
Figure~\ref{fig:3yr} reports coalition membership, daily and
cumulative coalition value over the three-year horizon. The dynamic
coalition overtakes the static grand coalition at $d=473$
and ends with $\PoS = 1.124$ and $\PoD = 1.057$ (94.6\% of the
full-information optimum).
\paragraph{Horizon comparison}
For seed 4, the dynamic/static ratio is $1.2385$ at one year
and $1.1242$ at three years. Because these runs are not pathwise
continuations, this comparison does not establish an asymptotic
trend. Across 30 seeds, the median ratios are $1.0876$ and $1.1327$;
\rev[R3.M3]{the first-to-third-quartile intervals, containing the middle half
of the runs, are \textcolor{black}{$[0.9945,1.1965]$} and \textcolor{black}{$[1.0688,1.2558]$}.
At one year, seed 4 gains \textcolor{black}{$23.9\%$}, above the third-quartile gain
of \textcolor{black}{$19.7\%$}: it shows a more favorable outcome than most runs.} The dynamic rule exceeds Grand in
22 of 30 one-year runs and 28 of 30 three-year runs. 

\begin{figure}[t]
\centering
\includegraphics[width=\columnwidth]{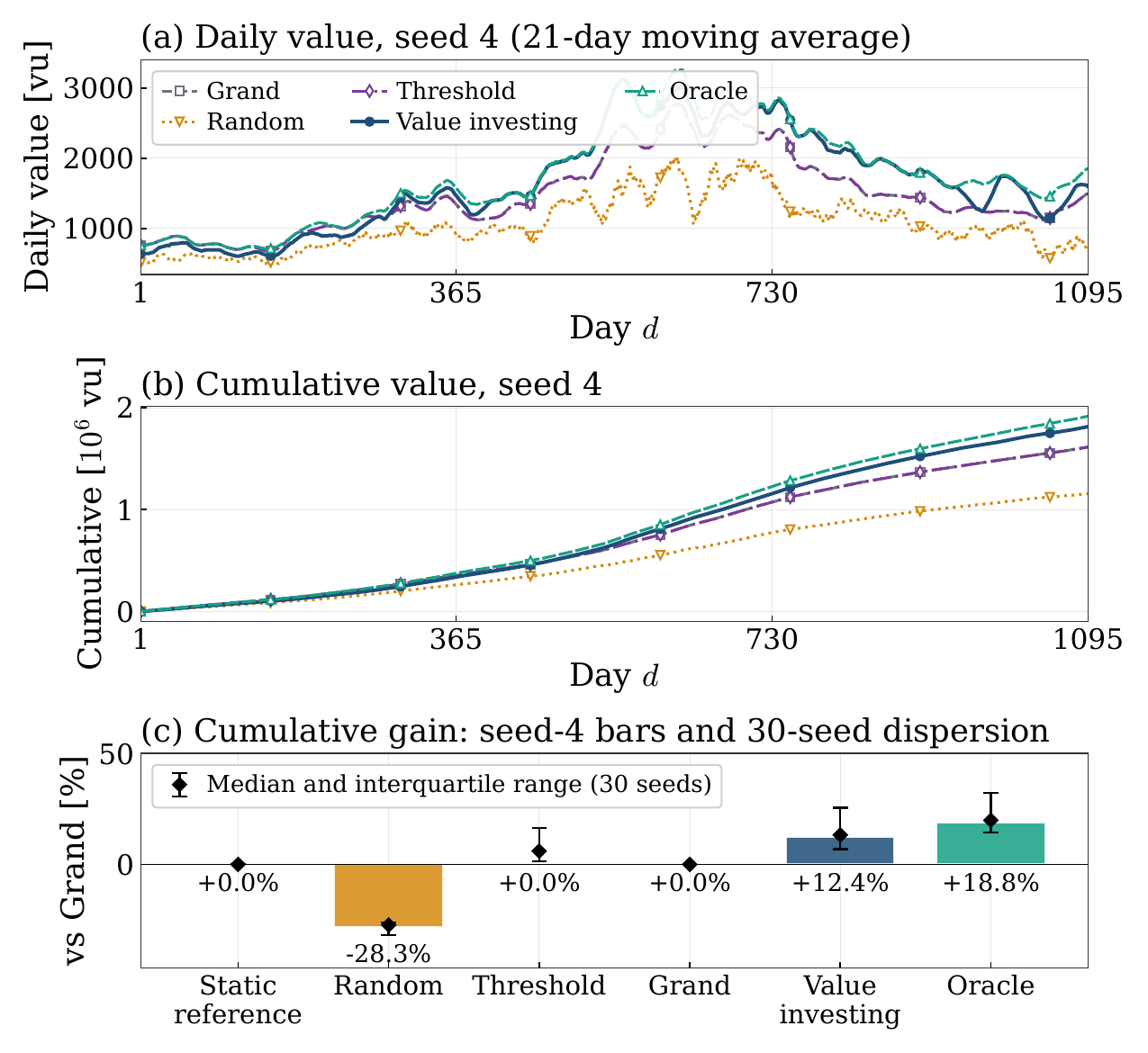}
\caption{Five strategies over three years. (a) Daily values, smoothed over
21 days. (b) Cumulative values. (c) Gain over Grand: seed-4 bars,
30-seed medians and interquartile ranges. The static reference equals Grand.
Smoothing affects display only; all totals use unsmoothed operational days.}
\label{fig:strategies}
\end{figure}

\begin{figure}[t!]
\centering
\includegraphics[width=\columnwidth]{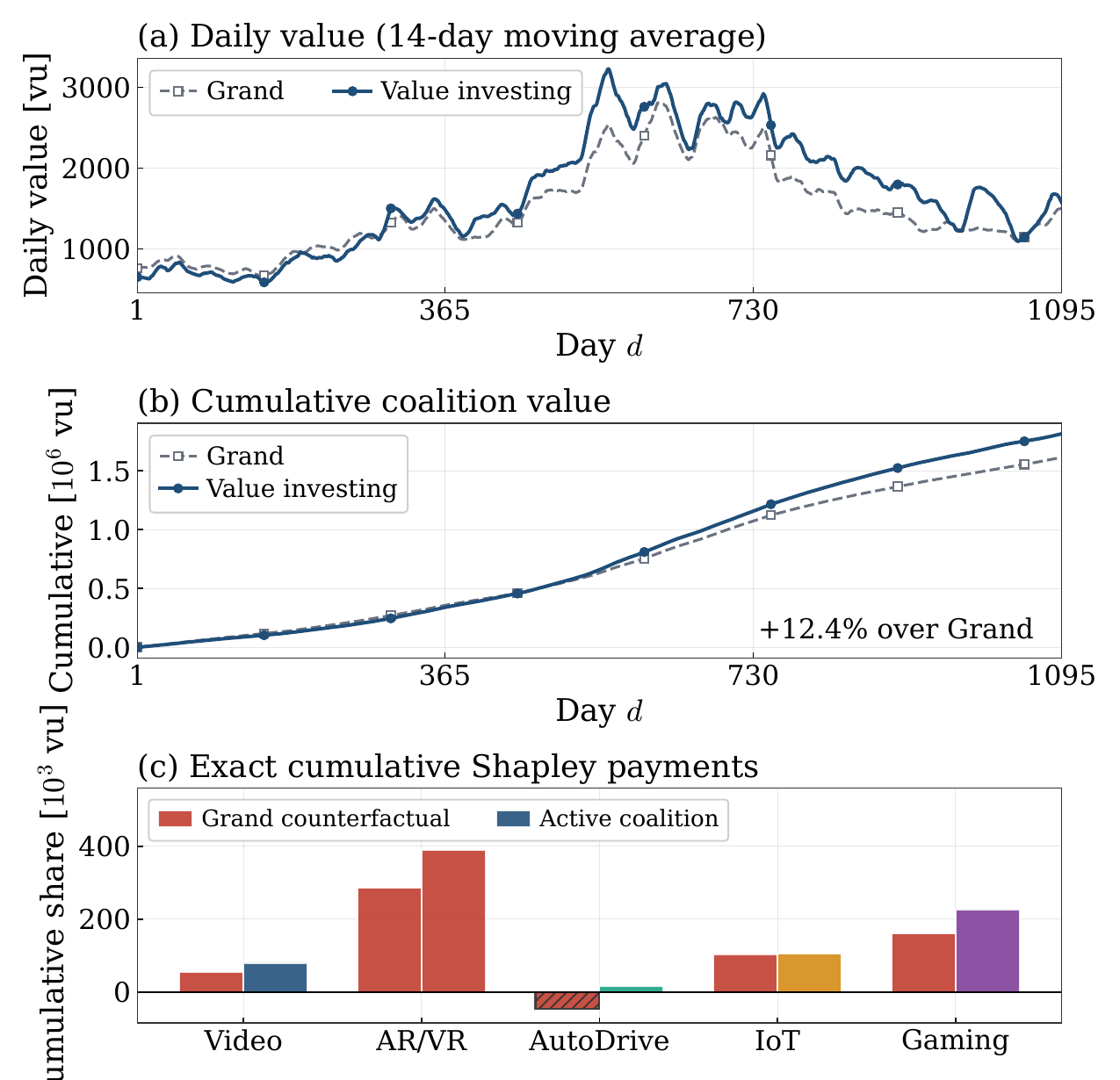}
\caption{Dynamic membership versus fixed Grand.
(a) Daily value, smoothed over 14 days. (b) Cumulative values, with a final $+12.4\%$ gain over Grand.
(c) Exact operational Shapley payments: Grand counterfactual and
realized active coalition, both in normalized value units.}
\label{fig:baseline}
\end{figure}

\paragraph{Strategy Comparison}
Figure~\ref{fig:strategies} compares five strategies over three years,
with seed-4 trajectories and dispersion across 30 seeds.
\rev[R3.M1c]{Reversing the update order reduces the median dynamic/static
ratio from \textcolor{black}{$1.133$} to \textcolor{black}{$0.686$}. Table~\ref{tab:results} and
Figure~\ref{fig:strategies} therefore evaluate the specified exit-first rule.} The
value-investing rule gains $12.4\%$ over Grand and captures $66.1\%$
of the oracle's available gain ($18.8\%$ over Grand).
\rev[ED.1]{The \emph{threshold-only} comparator admits SP~$i$ when
$X_i(d)>1.05$, using only its private profitability signal. It reproduces
Grand in the seed-4 three-year run (Table~\ref{tab:results}), illustrating
the limitation of ignoring displaced capacity.}
The marginal signal includes a capacity externality absent
from the individual profitability index. An active SP is coalitionally harmful when
$v_d(\mathcal S)-v_d(\mathcal S\setminus\{i\})<0$.
The individual index $X_i$ does not account for the capacity displaced
by its mandatory allocation. Consequently, individual profitability
alone does not generally determine whether participation increases
coalition value. The value-investing rule succeeds because its
decision variable is the marginal contribution $\Delta_i(d)$, which
prices the capacity externality through $\lambda^*(d)$, while the threshold in Definition~\ref{prop:theta} controls
admission conservatism; its effect is assessed by sensitivity tests. The random policy loses $28.3\%$ on this trajectory: the membership
selection rule matters for the observed gain.
\paragraph{Comparison with the Static Grand Coalition}
\label{subsec:baseline}
Figure~\ref{fig:baseline} compares the dynamic rule with fixed
grand-coalition membership under the same operational capacity model.
This isolates the effect of participation decisions. It is not an
independent numerical reproduction of the investment optimization
in~\cite{patane2023ccnc}.
The cumulative dynamic advantage becomes positive at day 473
and finishes at $12.4\%$; daily changes in the gap can have either sign.
Panel (c) distinguishes actual active-coalition payments from the
counterfactual payments under fixed grand-coalition membership.
Autonomous driving, whose reservation is 1500 mc, receives a cumulative
$+16{,}426$ vu in the realized dynamic coalition and would receive
$-45{,}844$ vu in the grand-coalition counterfactual. Its participation
rate is $20.2\%$, compared with $73.8\%$ for gaming and $74.1\%$
for AR/VR. The departure rule uses marginal contributions, so a
negative counterfactual Shapley payoff is an economic interpretation,
not the direct decision variable or a realized charge while inactive.
\paragraph{Operational Audit Calibration}
For each seed-4 operational day, the maximum NO payoff gain
over all 32 destination coalitions bounds every report-induced
membership outcome under verified settlement. The mean bound is
$\bar B=114.201$ vu and its maximum is $380.564$ vu.
Here $\bar B=D_{\mathrm{inv}}^{-1}\sum_{d=1}^{D_{\mathrm{inv}}}B_d$.
The sufficient penalty at day $d$ is denoted by
$\Pi_d^{\mathrm{suff}}(\alpha)=(1-\alpha)B_d/\alpha$.
With $c_a=22.5$ vu and the reference states generated at
$\alpha=0.15$, the design criterion gives $\alpha^*=0.9140$,
a mean sufficient penalty of $10.745$ vu, a maximum of $35.805$ vu,
and criterion value $21.489$ vu. This varies the audit parameter at
fixed reference states, rather than re-simulating membership at
$\alpha^*$. The daily penalty is state-dependent; its mean is not
a uniform guarantee.
For each tested audit/detection pair, the settlement check
covers $1095\times32=35{,}040$ destinations. Expected deviation
gains are nonpositive up to numerical tolerance under perfect or
$0.7$ detection probability, with the corresponding sufficient
penalty. These are conditional daily guarantees with no false
positives and full recovery; they do not establish repeated-game
truthfulness under delayed or erroneous audit feedback.
\begin{figure}[t]
\centering
\includegraphics[width=\columnwidth]{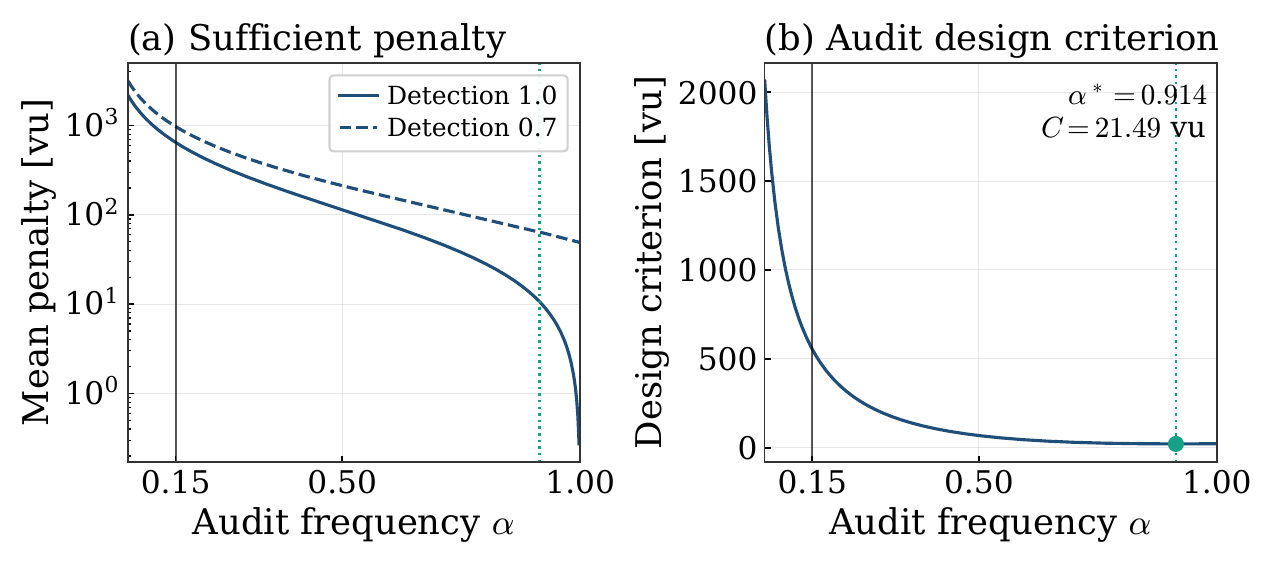}
\caption{Operational audit calibration at fixed seed-4 reference
states. (a) Mean sufficient daily penalty for detection probabilities
1 and 0.7. (b) The stated design criterion under perfect detection,
minimized at $\alpha^*=0.9140$, with value $21.489$ vu. Vertical lines mark $\alpha=0.15$ and $\alpha^*$.}
\label{fig:audit}
\end{figure}
\begin{table}[t]
\centering\small
\caption{Three-year cumulative values for seed 4 and ratios to Grand
across 30 seeds. Q1 and Q3 are the first and third quartiles.}
\label{tab:results}
\begin{tabular}{@{}lrr@{}}
\toprule
Strategy & Seed 4 [vu] & Ratio: median [Q1,Q3] \\
\midrule
Random & 1{,}156{,}990 & 0.726 [0.682,0.739] \\
Threshold & 1{,}614{,}270 & 1.060 [1.015,1.166] \\
Grand & 1{,}614{,}270 & 1.000 [1.000,1.000] \\
Value investing & 1{,}814{,}795 & 1.133 [1.069,1.256] \\
Oracle & 1{,}917{,}453 & 1.199 [1.143,1.322] \\
\bottomrule
\end{tabular}
\end{table}

\begin{figure}[t]
\centering
\includegraphics[width=\columnwidth]{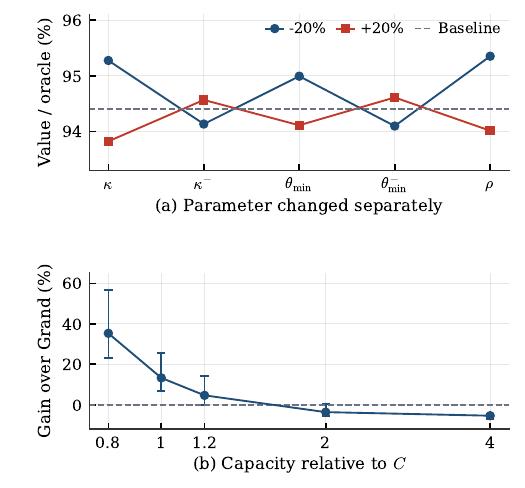}
\caption{Sensitivity across 30 three-year runs. (a) One policy parameter
varies at a time. (b) Capacity varies with the policy fixed.
Points show medians; bars in (b) contain the middle half of the runs.}
\label{fig:sensitivity}
\end{figure}

\paragraph{Robustness and computational checks}
\rev[R2.3.3,R4.M4]{Across 30 three-year seeds, varying each of the five
policy parameters by \textcolor{black}{$\pm20\%$} gives median dynamic/static ratios
from \textcolor{black}{$1.1227$} to \textcolor{black}{$1.1472$}. Figure~\ref{fig:sensitivity}(a) covers all
five parameters. Larger \textcolor{black}{$\kappa$} reduces participation and can
exclude beneficial services. Table~\ref{tab:sensitivity} reports
mean active SPs, daily endpoint changes (Churn), and all within-day
entries and exits (Events). Similar total value can conceal different
participation patterns.}
\begin{table}[t]
\centering\small
\caption{Participation sensitivity: median results at $-20\%$/$+20\%$
of each parameter; churn and events are averaged per day.}
\label{tab:sensitivity}
\setlength{\tabcolsep}{3pt}
\begin{tabular}{@{}lrrr@{}}
\toprule
Parameter & Active & Churn & Events\\
\midrule
Baseline & 2.48 & 0.123 & 3.22\\
$\kappa$ & 2.54/2.41 & 0.132/0.116 & 3.35/3.10\\
$\kappa^-$ & 2.61/2.42 & 0.126/0.129 & 3.65/3.05\\
$\theta_{\min}$ & 2.51/2.45 & 0.127/0.121 & 3.30/3.15\\
$\theta^-_{\min}$ & 2.56/2.41 & 0.127/0.126 & 3.50/3.05\\
$\rho$ & 2.35/2.70 & 0.127/0.133 & 2.83/4.00\\
\bottomrule
\end{tabular}
\end{table}
\rev[R3.M1b,R2.3.3]{At \textcolor{black}{$0.8C$}, median selection gain is \textcolor{black}{$35.3\%$};
at \textcolor{black}{$2C$}, selection loses \textcolor{black}{$3.7\%$} (Figure~\ref{fig:sensitivity}(b)).
Extra capacity reduces competition, while the admission margins
still exclude services. For \textcolor{black}{$\rho=0.04,0.05,0.06$}, median
\textcolor{black}{$\alpha^*$} is \textcolor{black}{$0.896,0.898,0.893$}. Each \textcolor{black}{$\rho$} regenerates coalition
states at nominal audit probability \textcolor{black}{$0.15$}; those states are then fixed
to evaluate the audit criterion. Thus \textcolor{black}{$\rho$} affects \textcolor{black}{$\alpha^*$} through
participation, not directly through its formula.}
\rev[R3.M2,R4.M1]{We also test standardized Student-\textcolor{black}{$t_3$} log-innovations,
compound-Poisson log-demand jumps (rate \textcolor{black}{$0.03$}/day, Gaussian jump
standard deviation \textcolor{black}{$0.35$}), weekly modulation, Zipf initial popularity
(exponent \textcolor{black}{$1.2$}), Pareto initial popularity (shape \textcolor{black}{$1.5$}), and 32
stationary active/idle (ON/OFF) sources per SP with Pareto durations
(shape \textcolor{black}{$1.4$}), following Taqqu et al.~\cite{taqqu1997onoff}.
The finite ON/OFF aggregate modulates the GRW; the aggregation limit
motivates persistence across scales, without establishing exact
self-similarity of the resulting demand. Respective median
dynamic/static ratios are \textcolor{black}{$1.1499$}, \textcolor{black}{$1.1996$}, \textcolor{black}{$1.1328$}, \textcolor{black}{$1.0411$},
\textcolor{black}{$1.0111$}, and \textcolor{black}{$1.1479$}. Changing initial service popularity reduces
the gain most. Student-\textcolor{black}{$t$} log-increments have no finite positive
exponential moment, so the GRW mean formula does not apply.}
\rev[R4.M5]{With a \textcolor{black}{$10\%$} false-positive audit rate and 30-day
additional feedback delay, median mean trust falls from \textcolor{black}{$0.9452$}
to \textcolor{black}{$0.9249$}, and the median dynamic/static ratio from \textcolor{black}{$1.1327$}
to \textcolor{black}{$1.1188$}. These tests concern trust and membership; Appendix~\ref{app:audit}
states separately the conditions for financial deterrence.}
Single-day synthetic Shapley tests use $M=50,100$ and 128 or 512
permutations. Their simultaneous Hoeffding radii can be too wide
to certify the sign of a small payoff; multi-year large-coalition
simulations remain future work.

\section{Conclusion}
\label{sec:concl}

We developed a dynamic value-investing coalitional game for edge
computing co-investment that simultaneously addresses determinism,
static membership, and full-information assumptions of prior work.
The two-stage decomposition separates the irreversible CAPEX from
the daily operational decisions; \rev[MR.1]{mandatory SLA allocations can produce negative operational
Shapley payoffs, whose implementation as fees requires contractual
settlement;} the margin-of-safety threshold, motivated by a benchmark uncertainty approximation, guides
participation under partial information; and the audit-and-trust
mechanism provides a conditional daily truthfulness guarantee for the NO.
\rev[ED.3]{Dynamic membership can improve operational value by releasing scarce
capacity from services whose reservations displace more valuable uses.
The gain depends on the update order and can disappear as capacity
becomes abundant.}

The most direct extensions for future work are: a
non-transferable-utility formulation, when SPs cannot freely
transfer payoffs; the integration of network delay into effective
deadlines and energy costs into the per-SP utility; the analysis of
$\PoS$ as $D_{\mathrm{inv}} \to \infty$; and a multi-NO
generalisation. Further work includes calibrated queueing and
latency models, calibration of investment participation constraints,
\rev[R2.3.5]{joint selection of \textcolor{black}{$\kappa$}, \textcolor{black}{$\theta_{\min}$} and \textcolor{black}{$\rho$}, general
equilibrium-convergence conditions, and multi-year simulations with
\textcolor{black}{$M=50$} or \textcolor{black}{$100$} SPs and competing or cooperating NOs.}

\appendices

\section{KKT Solution and Individual Value}
\label{app:kkt}
\begin{proof}[Proof of Proposition~\ref{prop:kkt}]
For a fixed day, we may suppress the day index in utility and multiplier notation.
By Lemma~\ref{lem:concave} the objective of \eqref{eq:char_function}
is strictly concave on a convex polyhedron, so the KKT conditions are necessary and sufficient~\cite{boyd2004}. With multipliers $\lambda \geq 0$ (capacity) and
$\nu_i \geq 0$ (SLA floor), stationarity reads
$\nu_i = \lambda - Y_i e^{-\xi h_i}$ with
$Y_i \triangleq \xi\beta_i \Lambda^i_d F = X_i(d)\,q$ by
\eqref{eq:Xi}. For an interior SP ($\nu_i = 0$) inversion gives
$h^*_i = \xi^{-1}\ln(X_i q/\lambda^*)$; an SLA-bound SP sits at
$h^*_i = \underline h_i$, which occurs iff
$Y_i e^{-\xi \underline h_i} \leq \lambda^*$. Together these give
\eqref{eq:hstar}. The capacity constraint binds because each $U^i_d$
is strictly increasing. Finally, let $H(\lambda)$ denote the total allocation requested at price
$\lambda$. Then $H(\lambda)\triangleq \sum_i \max\{\underline h_i,\,
\xi^{-1}\ln(X_i q/\lambda)\}$ is continuous and weakly decreasing,
with $H \to \infty$ as $\lambda \to 0^+$ and
$H\to\sum_i\underline h_i<C$ as $\lambda\to\infty$.
Continuity gives a solution; $H$ is strictly decreasing whenever
$H>\sum_i\underline h_i$, so the solution is unique.
\end{proof}

\begin{proof}[Derivation of \eqref{eq:vi}]
$v^i(d)$ is the value of $\max_{h \geq 0} [U^i_d(h) - qh]$, in which
SP~$i$ buys capacity at the reference value $q$ alone. For $X_i>1$, the first-order condition (FOC)
$Y_i e^{-\xi h} = q$ gives the individual optimizer $h^\circ_i=\xi^{-1}\ln X_i$ and
$e^{-\xi h^\circ_i} = 1/X_i$, so
\begin{equation*}
  U^i_d(h^\circ_i) - q h^\circ_i
  = \tfrac{q}{\xi}(X_i - 1) - \tfrac{q}{\xi}\ln X_i
  = \tfrac{q}{\xi}\,g(X_i),
\end{equation*}
using $Y_i = X_i q$. For $X_i\leq1$, $\partial_h U_i(0)=qX_i\leq q$;
concavity therefore gives the optimum $h=0$ and value zero.
Together these cases yield \eqref{eq:vi}.
\end{proof}

\section{Separability}
\label{app:sep}
\begin{proof}[Proof of Proposition~\ref{prop:sep}]
\emph{Knife-edge.} Under the interior-allocation assumptions of Proposition~\ref{prop:sep}, at $\lambda^*=q$ we have $h^*_i = \xi^{-1}\ln X_i$ for every
member and $U^i_d(h^*_i) = v^i(d) + q h^*_i$ by the derivation of
\eqref{eq:vi}. Summing over $\mathcal{P}(\mathcal{S})$ and subtracting
$q\sum_i h^*_i$ gives \eqref{eq:sep_exact}.

\emph{Interior allocation away from the knife-edge.} With
$e^{-\xi h^*_i} = \lambda^*/(X_i q)$ the same substitution yields
$U^i_d(h^*_i) - q h^*_i = v^i(d) - (q/\xi)\,g(\lambda^*/q)$, where the
remainder depends on $\mathcal{S}$ only through
$\lambda^*(\mathcal{S},d)$ and not on $i$. Hence
\begin{equation}
  \tilde v_d(\mathcal{S}) =
   \!\!\sum_{i \in \mathcal{P}(\mathcal{S})}\!\! v^i(d)
   - \frac{|\mathcal{P}(\mathcal{S})|\,q}{\xi}\,
     g\!\bigl(\lambda^*(\mathcal{S},d)/q\bigr).
  \label{eq:sep_offedge}
\end{equation}
Since $g'(1) = 0$ and $g''(1) = 1$, expanding $g$ about $u = 1$ gives
$g(\lambda^*/q) = \tfrac12(1-\lambda^*/q)^2 + O((1-\lambda^*/q)^3)$,
hence the explicit bound

\begin{equation}
  \Bigl|\tilde v_d(\mathcal{S}) -
   \!\!\sum_{i \in \mathcal{P}(\mathcal{S})}\!\! v^i(d)\Bigr|
  = \frac{|\mathcal{P}(\mathcal{S})|}{2\,\xi\,q}
    \bigl(q - \lambda^*\bigr)^2 + O\bigl((q-\lambda^*)^3\bigr),
  \label{eq:sep_bound}
\end{equation}

When an SLA floor binds, the interior expression
must be replaced by the corresponding boundary
allocation. The comparison between $\lambda^*$ and $q$
alone does not determine whether this occurs.
\end{proof}

\section{Conditional Submodularity}
\label{app:convex}
\begin{proof}[Proof of Theorem~\ref{thm:convex}]
{
Suppress the day index. Let $W(\mathcal P,c)$ denote the optimal
utility of SP set $\mathcal P$ at capacity $c$, and
$\lambda_{\mathcal P}(c)$ its capacity multiplier.
Let $H_{\mathcal P}(\lambda)$ be total requested allocation at price
$\lambda$. For SP sets $\mathcal A\subseteq\mathcal B$, define
\[
H_{\mathcal P}(\lambda)=\sum_{i\in\mathcal P}
\max\left\{\underline h_i,\frac1\xi\ln\frac{X_iq}{\lambda}\right\}.
\]
Since $H_{\mathcal A}(\lambda)\leq H_{\mathcal B}(\lambda)$,
capacity clearing gives
$\lambda_{\mathcal A}(c)\leq\lambda_{\mathcal B}(c)$ for
$c>\sum_{i\in\mathcal B}\underline h_i$; set
$\lambda_{\emptyset}(c)=0$.
For $j\notin\mathcal B$, let
$\mathcal H_{\mathcal P}=[\underline h_j,C-\sum_{i\in\mathcal P}\underline h_i]$
be its feasible allocation interval.
The marginal contribution of $j$ is
\[
\max_{h\in \mathcal H_{\mathcal P}}
\left\{U_j(h)-\int_{C-h}^{C}\lambda_{\mathcal P}(c)\,dc\right\},
\]
by the envelope theorem. Since
$\mathcal H_{\mathcal B}\subseteq \mathcal H_{\mathcal A}$ and the integrated shadow
price is no smaller for $\mathcal B$, the maximum for $\mathcal B$
cannot exceed that for $\mathcal A$. This proves submodularity.
}
\end{proof}

\section{Benchmark and Operational Shapley Values}
\label{app:shapley}
\begin{proof}[Proof of Theorem~\ref{thm:shapley}]
For the benchmark $\bar v_d$, SP~$i$ has marginal contribution
$v^i(d)$ exactly when the NO precedes it. This occurs in half the
permutations, proving \eqref{eq:shapley_A}; efficiency gives the NO's
share. Each nonnegative $v^i(d)$ weights the unanimity game on
$\{i,\NO\}$, so their sum is supermodular and has a nonempty
core~\cite{shapley1971}.
For the operational game $v_d$, the general Shapley formula
gives \eqref{eq:shapley_B} in both capacity regimes. Each predecessor
coalition is evaluated using \eqref{eq:char_function}; coalition
values without the NO are zero.
\end{proof}

\section{Strategy-Proofness and Cost-Optimal Audit}
\label{app:audit}
\rev[R4.M5]{\begin{proof}[Proof of Theorem~\ref{thm:sp}]
Let \textcolor{black}{$G(\mathbf r)$} be the NO's payoff change when a false declaration \textcolor{black}{$\mathbf r$} goes undetected. An audit occurs with probability
\textcolor{black}{$\alpha$} and then voids the inflated outcome and levies \textcolor{black}{$\Pi$}, so \(
  \mathbb{E}[\varphi_{\NO} \mid \mathbf r] - \varphi_{\NO}
   = (1-\alpha)\,G(\mathbf r) - \alpha\,\Pi .\)
Under the daily-payoff assumptions of Theorem~\ref{thm:sp},
truth-telling is weakly optimal iff this is non-positive for every
admissible \textcolor{black}{$\mathbf r$}, which is \eqref{eq:SP_condition}. For a feasible expansion without departures in the separable
benchmark, Theorem~\ref{thm:shapley}(i) gives
\textcolor{black}{$G^+(\mathbf r)=\frac12\sum_{i\in\mathcal J(\mathbf r)}v^i(d)$}.
This expression does not replace the operational gain in general.
\textbf{Audit errors and delay.} Let \textcolor{black}{$p_{\mathrm{det}}>0$} be detection
probability conditional on auditing, and \textcolor{black}{$b_{\mathrm{rec}}\in(0,1]$}
the discount applied to recovery and penalty collection. With no false
positives, expected deviation gain is
\textcolor{black}{$(1-\alpha p_{\mathrm{det}}b_{\mathrm{rec}})G(\mathbf r)
-\alpha p_{\mathrm{det}}b_{\mathrm{rec}}\Pi$}.
Replacing \textcolor{black}{$G$} by \textcolor{black}{$B_d$} gives the sufficient penalty
\textcolor{black}{$\Pi\geq(1-\alpha p_{\mathrm{det}}b_{\mathrm{rec}})B_d/
(\alpha p_{\mathrm{det}}b_{\mathrm{rec}})$}.
Without recovery, the numerator is \textcolor{black}{$B_d$}.
False positives require comparison with the truthful report's sanction;
observation errors that corrupt the reference values also affect \textcolor{black}{$B_d$}.
Pending audits do not update trust until their results arrive.
These statements concern daily financial incentives, not continuation payoffs.
\end{proof}

}\begin{proof}[Proof of Corollary~\ref{cor:opt_audit}]
Use the operational maximum in \eqref{eq:Pi0star} and set
the penalty at \eqref{eq:Pistar}. The following minimization concerns
the stated design criterion, not the expected penalties paid under
truthful play. Writing
$E \triangleq \mathbb{E}[\Pi^*_0(d)]$ and setting $\Pi$ at that
threshold, the design-criterion value is
\begin{equation}
  C_{\mathrm{audit}}(\alpha)
   = \alpha\,c_a + (1-\alpha)\,\tfrac{1-\alpha}{\alpha}\,E
   = \alpha\,c_a + \tfrac{(1-\alpha)^2}{\alpha}\,E ,
  \label{eq:app_Caudit}
\end{equation}
with $C_{\mathrm{audit}}'(\alpha) = c_a - E\,(1-\alpha^2)/\alpha^2$
and $C_{\mathrm{audit}}''(\alpha) = 2E/\alpha^3 > 0$. The map is
strictly convex, so the stationary point is the unique minimiser:
$c_a\alpha^2 = E(1-\alpha^2)$ gives \eqref{eq:alphastar}. The same minimization applies to a fixed reference distribution
of $B_d$. Varying the audit frequency may also change future trust
and coalition membership, so this calculation is a frozen-state
design criterion rather than a repeated-game cost optimum.
\end{proof}
\section*{Acknowledgment}
\rev[AI]{The author used ChatGPT (OpenAI) for language editing and
\LaTeX{} preparation throughout the manuscript, for drafting the
clarifications in Sections~\ref{sec:model} and~\ref{sec:shapley}, and
for assistance with simulation and figure-generation code and numerical
consistency checks supporting Section~\ref{sec:results}. The author
retains full responsibility for the manuscript's scientific content
and final presentation.}

\end{document}


\maketitle

This supplement accompanies the main paper~\cite{main}. It gives
additional details on traffic, Shapley sampling, the separable benchmark,
ordered coalition updates, and numerical results. The main proofs remain
in the manuscript appendices. Symbols have the meanings used in the main
paper; equations, propositions, tables and figures in this document are
numbered independently.

\section{GRW Traffic Process: Justification}
\label{supp:grw}
\paragraph{Multiplicative growth}
The Geometric Random Walk (GRW) provides a positive demand model in which
relative changes compound. It represents evolving service popularity,
without requiring an additive Gaussian model to remain positive.

\paragraph{Log-normality and conditional moments}
The model specifies
\[
 \ln\Lambda_d^i=\ln\Lambda_0^i+
 (\mu_i-\sigma_i^2/2)d+\sigma_i\sum_{k=1}^d\epsilon_k^i,
 \qquad \epsilon_k^i\sim\mathrm{N}(0,1),
\]
with independent innovations across services and days. Hence
$\mathbb E[\Lambda_d^i]=\Lambda_0^i e^{\mu_i d}$, whereas the
expected log increment per day is $\mu_i-\sigma_i^2/2$.
For a residual horizon $r$, the conditional coefficient of variation is
$\sqrt{e^{\sigma_i^2r}-1}$. Its approximation $\sigma_i\sqrt r$
requires $\sigma_i^2r\ll1$ and motivates the admission margin; it is
not an exact risk bound for cumulative operational Shapley payments.

\paragraph{Applicability}
Independent Gaussian log increments do not describe bursts, periodicity
or long-range dependence. Demand trajectories can diverge across services,
changing the capacity displaced by their reservations. This can favor
membership updates, but does not imply that their advantage grows
monotonically with the horizon. The one-year and three-year runs use
horizon-dependent random arrays and are not pathwise continuations.

\section{Shapley Value Scalability via Sampling}
\label{supp:sampling}
The exact five-SP implementation caches all $2^M$ coalition values and
computes operational Shapley allocations, including the NO as a veto
player. A single active-game allocation uses $O(M2^M)$ weighted-subset
operations. The full payoff table over all active SP sets, needed for
the exhaustive operational audit bound, uses $O(M3^M)$ such operations.

\paragraph{Permutation sampling}
For larger active games, draw $K$ independent uniform player permutations
$\pi_1,\ldots,\pi_K$ and use~\cite{castro2009shapley}
\begin{equation}
 \hat\varphi_i^-(d)=\frac1K\sum_{k=1}^K
 \bigl[v_d(\mathcal S_i^{\pi_k}\cup\{i\})-
 v_d(\mathcal S_i^{\pi_k})\bigr].
 \label{eq:supp:castro}
\end{equation}
Here $\mathcal S_i^{\pi_k}$ contains the predecessors of $i$.
The estimator is unbiased. With
$R=\sum_j\beta_j\Lambda_d^jF$, each marginal lies in $[-R,R]$.
Hoeffding's inequality and a union bound give simultaneous additive
accuracy $\delta_{\mathrm{err}}$ for all $M+1$ players with failure
probability at most $\eta_{\mathrm{fail}}$ if
\begin{equation}
 K\geq\frac{2R^2}{\delta_{\mathrm{err}}^2}
 \log\frac{2(M+1)}{\eta_{\mathrm{fail}}}.
\end{equation}
Relative-error requirements become ill-conditioned near a zero payoff;
the implementation therefore reports an additive confidence radius.
A negative estimated payoff is not sign-certified if its interval
contains zero. Even a certified negative operational payoff is a
theoretical allocation signal; implementing it as a monetary fee requires
contractual payment, enforcement and recovery terms.

\paragraph{Stratified sampling}
Permutation designs can reduce sampling variance~\cite{mitchell2022sampling}.
For example, conditioning on a player's rank gives equally weighted
strata, each of which can be sampled uniformly. A variance reduction
factor is not assumed here. Capacity multipliers are monotone under set
inclusion, not under cardinality comparisons between arbitrary coalitions.

\paragraph{Decision and audit calculations}
Participation uses entry and departure marginals, rather than sampled
Shapley payoffs. The audit covers both the declared coalition value and
the marginal-signal vector. Sampling a fixed game's payoffs does not by
itself certify the maximum deviation gain over all destination coalitions.
If approximate payoffs are used in an audit bound, their uncertainty and
the coverage of candidate destinations must also be controlled.

\section{Expected Separable-Benchmark Shapley Payoff}
\label{supp:Eshapley}
The benchmark payoff is
$\varphi_i^+(d)=\frac{q}{2\xi}g(X_i(d))\mathbf1_{\{X_i(d)>1\}}$,
where $g(x)=x-1-\ln x$ and $X_i(d)=A_i\Lambda_d^i$.
The truncation at $X_i=1$ follows from the nonnegative individual
capacity choice. This benchmark is distinct from the operational game
in either capacity regime.

\begin{proposition}[Expected benchmark payoff]
Let $m_i=\ln(A_i\Lambda_0^i)+(\mu_i-\sigma_i^2/2)d$ and
$s_i=\sigma_i\sqrt d>0$. With $\Phi_N$ and $\phi_N$ denoting the
standard normal distribution and density,
\begin{equation}
\begin{split}
\mathbb E[\varphi_i^+(d)]
=\frac{q}{2\xi}\biggl[
&e^{m_i+s_i^2/2}\Phi_N\!\left(\frac{m_i+s_i^2}{s_i}\right)\\
&-(1+m_i)\Phi_N\!\left(\frac{m_i}{s_i}\right)
-s_i\phi_N\!\left(\frac{m_i}{s_i}\right)\biggr].
\end{split}
\label{eq:supp:Ephi}
\end{equation}
For $s_i=0$, the expectation is the deterministic benchmark payoff.
\end{proposition}
\begin{proof}
Write $Y=\ln X_i(d)\sim\mathrm N(m_i,s_i^2)$. Integrating the
normal density over $Y>0$ gives
\begin{align*}
\mathbb E[e^Y\mathbf1_{\{Y>0\}}]
 &=e^{m_i+s_i^2/2}\Phi_N((m_i+s_i^2)/s_i),\\
\Pr(Y>0)&=\Phi_N(m_i/s_i),\\
\mathbb E[Y\mathbf1_{\{Y>0\}}]
 &=m_i\Phi_N(m_i/s_i)+s_i\phi_N(m_i/s_i).
\end{align*}
Combining these three terms yields \eqref{eq:supp:Ephi}.
\end{proof}
\begin{corollary}[Benchmark growth]
For $\mu_i>0$, the expected benchmark payoff is asymptotic to
$(q/(2\xi))A_i\Lambda_0^i e^{\mu_i d}$ as $d\to\infty$.
\end{corollary}
\begin{proof}
The leading truncated exponential moment is asymptotic to
$A_i\Lambda_0^i e^{\mu_i d}$, since
$(m_i+s_i^2)/s_i\to\infty$. The remaining terms grow at most linearly
in $d$. This statement concerns the benchmark, not an asymptotic
dynamic/static value ratio.
\end{proof}

\section{Observed Action Times}
\label{supp:stopping}
The first observed join or leave time is the first operational day on
which the ordered sweep executes that action. An action not observed by
$D_{\mathrm{inv}}$ is recorded as $\infty$. Traffic drift alone does not
establish eventual entry or departure: coalition values, competing
services, trust and thresholds also enter the decision. No almost-sure
eventual-action claim is made. An SP can leave and re-enter within one
day, so action counts and changes in daily endpoint membership differ.

\section{Ordered Updates and Convergence Statistics}
\label{supp:equil}
Each day starts from the preceding day's endpoint. Exits are processed
before entries, in the order Video, AR/VR, AutoDrive, IoT, Gaming, with
values recomputed after every change. Reported daily values use one
sweep. Repeated frozen-day sweeps are a separate diagnostic.

For fixed traffic, trust and thresholds, the deterministic sweep acts
on $2^M$ SP subsets. Among $2^M+1$ consecutive endpoints, two must
coincide, giving a periodic orbit. Equal consecutive endpoints do not
exclude intermediate exit and re-entry and therefore do not prove an
individual equilibrium. Selecting one state from a cycle would not
remove its eligible moves.

A sufficient convergence condition is given in the main paper: on the
reachable feasible sets $\mathcal K$, each player's largest eligible
departure marginal $L_i$ is smaller than its smallest eligible entry
marginal $J_i$. Choosing $L_i<a_i<J_i$ makes
$w_d(\mathcal P)-\sum_{i\in\mathcal P}a_i$ strictly increase at every
eligible move. No set then repeats after a move, so at most
$|\mathcal K|-1$ membership changes occur before equilibrium.

\begin{table}[t]
\centering\small
\caption{Frozen-day diagnostics: seed 4, 1095 operational days.}
\label{tab:supp:convergence}
\begin{tabular}{@{}lr@{}}
\toprule
Diagnostic & Result\\\midrule
Period-one sweep orbits & 1093 days\\
Period-two sweep orbits & 2 days\\
Mean transitions to first repetition & 1.128\\
Maximum transitions to first repetition & 3\\
Daily endpoints satisfying individual equilibrium & 97 days\\
Days with an equilibrium in exact enumeration & 97 days\\
Days satisfying the reachable-set potential condition & 97 days\\
\bottomrule
\end{tabular}
\end{table}
Table~\ref{tab:supp:convergence} shows that repetition is usually rapid,
while individual equilibrium is uncommon. On the remaining 998 days,
exact enumeration finds no individual equilibrium. These observations
do not establish convergence for arbitrary service populations.

\section{Numerical Setup, Parameters, and Additional Figures}
\label{supp:numerical}
\subsection{Parameters and Scope}
The fixed capacity $C$ is the installed CPU budget determined in Stage~I.
The framework targets long-term infrastructure co-investment and daily
service-coalition management within that budget. Dynamic scaling,
elastic cloud offloading and joint orchestration of heterogeneous
CPU, memory and bandwidth resources lie outside this model.
SLA floors are contractual reservations supplied as inputs; they are
not inferred from queueing, throughput or end-to-end latency guarantees.
Utilities, payoffs, penalties and audit costs use normalized value units
(vu), not dollars. The reference marginal value $q$ is distinct from
the deployment cost $q_{\mathrm{cap}}$.

\begin{table}[t]
\centering\small
\caption{Operational parameters. Capacity is rounded for display.}
\label{tab:supp:params}
\setlength{\tabcolsep}{4pt}
\begin{tabular}{@{}lll@{}}
\toprule
Symbol & Value & Meaning / unit\\\midrule
$M$ & 5 & Service providers\\
$D_{\mathrm{inv}}$ & 365, 1095 & Horizon [day]\\
$T$, $\Delta t$ & 96, 900 & Slots; slot duration [s]\\
$F$ & 96 & Dimensionless profile sum\\
$C$ & 3100.394 & CPU capacity [mc]\\
$q$ & 0.10 & Reference value [vu/mc]\\
$\xi$ & $10^{-3}$ & Saturation [mc$^{-1}$]\\
$\theta_{\min}$, $\theta^-_{\min}$ & 0.050, 0.025 & Base margins\\
$\kappa$, $\kappa^-$ & 0.10, 0.05 & Volatility coefficients\\
$\alpha$, $\rho$ & 0.15, 0.05 & Audit; trust parameters\\
$c_a$ & 22.5 & Cost per audit [vu]\\
$\varepsilon$ & $-15\%$ & Scalar report test\\
Seeds & $0,\ldots,29$ & Reference trajectory: 4\\
\bottomrule
\end{tabular}
\end{table}

\begin{table*}[t]
\centering\small
\caption{Service-class inputs. The coefficients $\beta_i$ convert the
traffic scale to normalized utility. Reservations are exogenous.}
\label{tab:supp:services}
\begin{tabular}{@{}lrrrrr@{}}
\toprule
Service & $\beta_i$ [vu/(req/s)] & $\Lambda_0^i$ [req/s]
 & $\mu_i$ [day$^{-1}$] & $\sigma_i$ [day$^{-1/2}$]
 & $\underline h_i$ [mc]\\\midrule
Video & 0.008 & 400 & 0.0008 & 0.040 & 30\\
AR/VR & 0.015 & 250 & 0.0012 & 0.060 & 300\\
AutoDrive & 0.025 & 150 & 0.0005 & 0.050 & 1500\\
IoT & 0.004 & 800 & 0.0006 & 0.030 & 0.3\\
Gaming & 0.012 & 350 & 0.0010 & 0.045 & 112.5\\
\bottomrule
\end{tabular}
\end{table*}

\begin{table*}[t]
\centering\small
\caption{Policy sensitivity: medians over 30 three-year runs.
Paired entries correspond to $-20\%$/$+20\%$ of one parameter;
the other parameters remain fixed.}
\label{tab:supp:sensitivity}
\begin{tabular}{@{}lrrrr@{}}
\toprule
Parameter & VI/Grand & Active & Churn/day & Events/day \\
\midrule
Baseline & 1.133 & 2.478 & 0.123 & 3.215 \\
$\kappa$ & 1.143/1.126 & 2.537/2.409 & 0.132/0.116 & 3.351/3.095 \\
$\kappa^-$ & 1.125/1.145 & 2.605/2.416 & 0.126/0.129 & 3.652/3.047 \\
$\theta_{\min}$ & 1.137/1.131 & 2.508/2.445 & 0.127/0.121 & 3.301/3.151 \\
$\theta^-_{\min}$ & 1.125/1.145 & 2.558/2.414 & 0.127/0.126 & 3.500/3.046 \\
$\rho$ & 1.147/1.123 & 2.347/2.700 & 0.127/0.133 & 2.832/4.000 \\
\bottomrule
\end{tabular}
\end{table*}

The aggregate reservation is 1942.8 mc, or 62.7\% of capacity.
AutoDrive has the largest reservation and therefore can displace
substantial capacity even when its current demand is low. Negative
Shapley payoffs reflect permutation-averaged marginal impact, whereas
the departure rule evaluates the marginal at the current decision point.

\subsection{Simulation Protocol and Sensitivity}
All totals exclude initialization day zero. Within a seed and horizon,
strategies share an SP-major demand-innovation array. The one-year and
three-year runs use different horizon-dependent arrays. Audit feedback
is available from the next day; initial trust is $1-\sqrt\rho$.
The baseline and sensitivity campaigns use 30 seeds. Churn counts
changes between daily endpoint sets; Events counts all within-day
entries and exits. Both are averaged per day before taking seed medians.

The policy changes in Table~\ref{tab:supp:sensitivity} have modest
effects on aggregate value but different effects on participation and
within-day activity. Capacity matters more: the median gain is 35.3\%
at $0.8C$, while selection loses 3.7\% at $2C$. With abundant capacity,
admission margins can exclude services that would add value. Reversing
the update order also substantially reduces performance; the results
therefore refer to the specified exit-first rule.

\paragraph{Operational audit calibration}
The all-destination bound compares the NO's verified payoff with the
truthful endpoint across all 32 feasible SP subsets. For seed 4, its
mean is 114.201 vu and maximum is 380.564 vu. At fixed reference states,
the stated criterion gives $\alpha^*=0.9140$ and mean sufficient penalty
10.745 vu. The penalty is state dependent; its mean is not a uniform
guarantee. These calculations assume enforceable settlement and rollback.
False-positive and delayed-feedback experiments assess trust and
membership separately from financial deterrence.

\paragraph{Alternative demand and reproducibility}
The accompanying code specifies Student-$t_3$, jump, weekly, Zipf,
Pareto and ON/OFF variants. The baseline, Student-$t_3$ and weekly
summaries reproduce the main paper. The supplied implementations of
jumps, Zipf and Pareto initial popularity, and ON/OFF modulation use
explicit random-stream and initialization conventions but do not
reproduce the four corresponding summaries in the main manuscript.
The numerical comparison and protocol definitions are provided in the
simulator's \path{outputs/PAPER_COMPARISON.md} report.

\subsection{Additional Figures}
Figures~\ref{fig:supp:membership}--\ref{fig:supp:misreport} use the
same operational model as the main paper. Curves marked as smoothed
are smoothed only for display; totals use unsmoothed daily values.

\section{Acronyms}
The acronym glossary is given in Table~I of the main paper.

\section*{Acknowledgment}
The author used ChatGPT (OpenAI) for language and \LaTeX{} preparation,
for assistance with the derivations and consistency checks in this
supplement, and for simulation and figure-generation code. The author
retains full responsibility for the scientific content and final presentation.

\clearpage
\onecolumn
\section*{Additional Figures}
\begin{supppanel}
\centering\includegraphics[width=\linewidth]{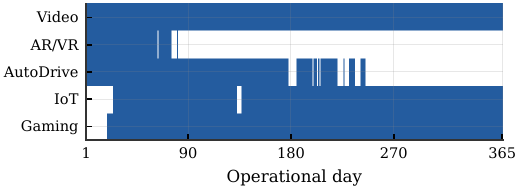}
\caption{One-year membership, seed 4. Operational suspension preserves
the investment rights described in the main paper.}
\label{fig:supp:membership}
\end{supppanel}\hfill
\begin{supppanel}
\centering\includegraphics[width=\linewidth]{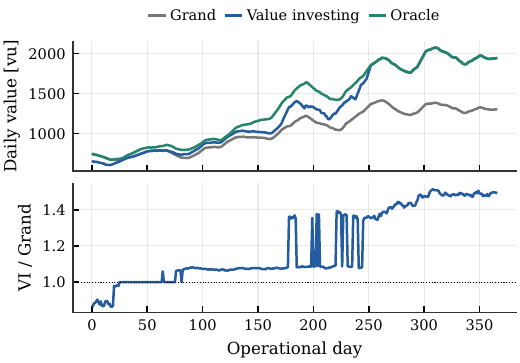}
\caption{One-year daily value (14-day display mean) and the dynamic/static
ratio. The reference gain is 23.9\%; it is not a time prefix of the
three-year experiment.}
\label{fig:supp:1yr}
\end{supppanel}
\par\medskip
\begin{supppanel}
\centering\includegraphics[width=\linewidth]{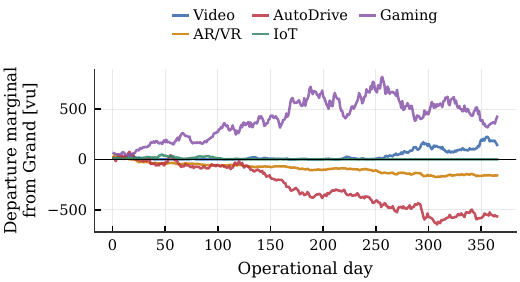}
\caption{One-year departure marginals from Grand. A negative marginal
means removing that SP increases coalition value at the current state;
it is distinct from a Shapley payoff and from a monetary charge.}
\label{fig:supp:marginals}
\end{supppanel}\hfill
\begin{supppanel}
\centering\includegraphics[width=\linewidth]{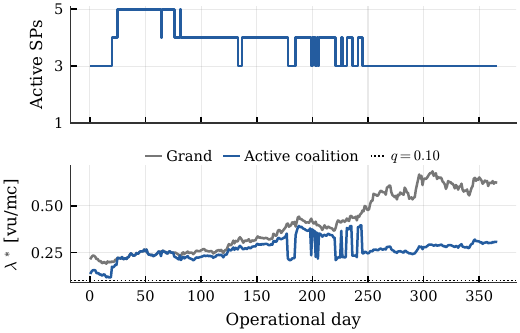}
\caption{One-year active-SP count and capacity shadow prices.
The active coalition and Grand can have different scarcity levels.}
\label{fig:supp:regime}
\end{supppanel}
\par\medskip
\begin{supppanel}[\textwidth]
\centering\includegraphics[width=\linewidth]{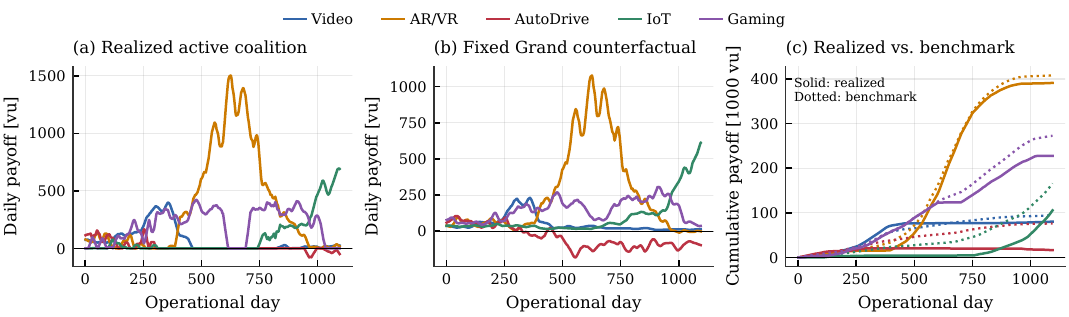}
\caption{Three-year Shapley trajectories, seed 4. (a) Realized active-coalition payoffs, zero while inactive.
(b) Fixed-Grand counterfactual payoffs. Daily curves use a 21-day display
mean. (c) Cumulative realized payoffs (solid) and separable-benchmark
payoffs (dotted). The benchmark uses a different coalition-value model;
these curves do not establish a universal ordering between allocations.
A negative operational payoff is an allocation signal whose implementation
as a fee requires a contract.}
\label{fig:supp:shapley}
\end{supppanel}
\clearpage
\begin{supppanel}
\centering\includegraphics[width=\linewidth]{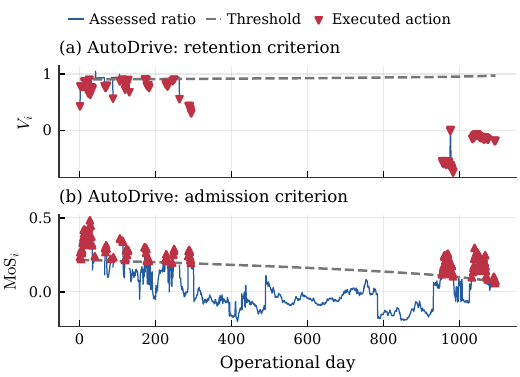}
\caption{AutoDrive's ratios at the actual exit and entry decision points.
Each panel uses its own threshold. The distinct scales do not establish
a hysteresis band; within-day exit and re-entry remain possible.}
\label{fig:supp:vratio}
\end{supppanel}\hfill
\begin{supppanel}
\centering\includegraphics[width=\linewidth]{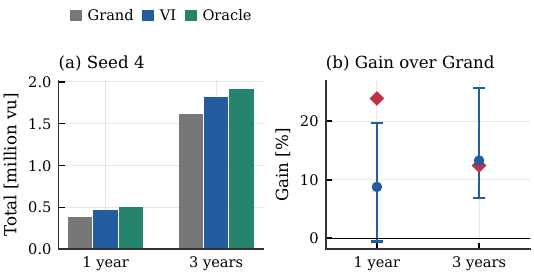}
\caption{Horizon comparison. (a) Seed-4 totals. (b) Dynamic/static gain:
medians and interquartile ranges across 30 seeds, with seed 4 marked by
a diamond. These runs do not establish an asymptotic gain trend.}
\label{fig:supp:summary}
\end{supppanel}
\par\medskip
\begin{supppanel}
\centering\includegraphics[width=\linewidth]{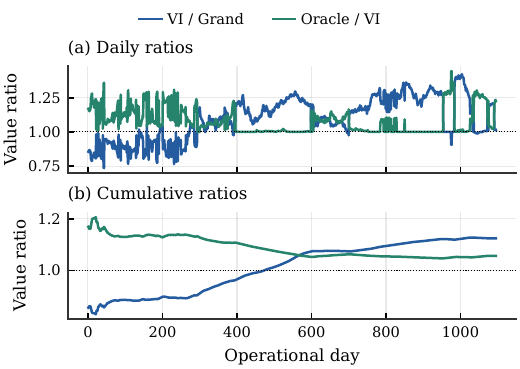}
\caption{Three-year daily and cumulative ratios. The final dynamic/static
ratio is 1.124 and $\PoD=1.057$, corresponding to 94.6\% of the
same-day oracle value. Neither cumulative gap nor daily ratio is monotone.}
\label{fig:supp:podpos}
\end{supppanel}\hfill
\begin{supppanel}
\centering\includegraphics[width=\linewidth]{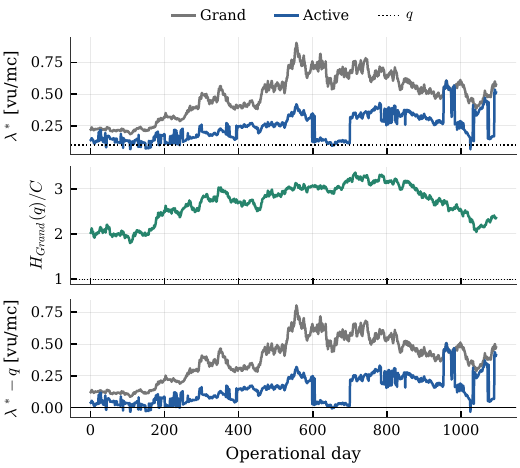}
\caption{Three-year scarcity diagnostics. Grand's shadow price exceeds
$q$ throughout; the active coalition's does so on 96.9\% of days.
The middle panel measures capacity requested by Grand at reference
value $q$ relative to installed capacity.}
\label{fig:supp:scarcity}
\end{supppanel}
\par\medskip
\begin{supppanel}[\textwidth]
\centering\includegraphics[width=\linewidth]{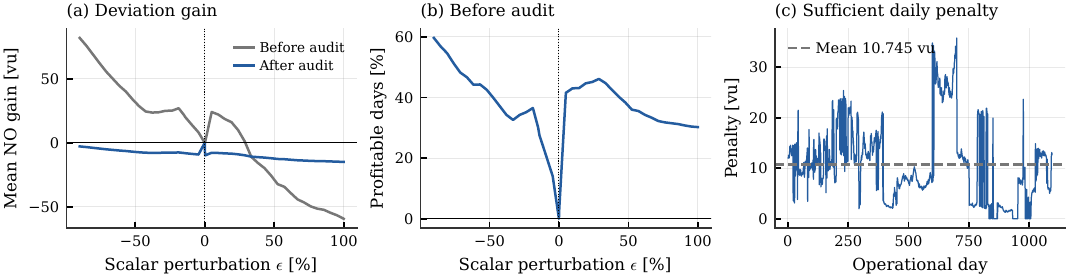}
\caption{Scalar declaration tests at fixed truthful states and trust.
Marginal signals remain truthful. (a) Mean NO deviation gain before and
after auditing. (b) Fraction of days on which a scalar perturbation is
profitable before audit. (c) State-dependent sufficient daily penalties
at $\alpha^*=0.9140$. These tests concern current-day verified settlement,
not repeated-game incentives.}
\label{fig:supp:misreport}
\end{supppanel}